\documentclass[lettersize,journal]{IEEEtran}

\usepackage[utf8]{inputenc}
\usepackage[T1]{fontenc}
\usepackage{url}
\usepackage{ifthen}
\usepackage{cite}
\usepackage[cmex10]{amsmath}
\usepackage{graphicx,psfrag,epsfig,epsf,latexsym,hhline,amssymb,multirow}
\usepackage[caption=false,font=normalsize,labelfont=sf,textfont=sf]{subfig}
\usepackage{pst-plot}
\usepackage{pstricks-add}
\usepackage{pifont}
\usepackage{graphicx}
\usepackage{amsthm}
\usepackage[linesnumbered,ruled,vlined]{algorithm2e}
\usepackage{array}
\usepackage{blindtext}
\usepackage{etoolbox}
\usepackage{comment}
\usepackage{epstopdf}
\usepackage{hyperref}
\usepackage{balance}
\hypersetup{
    bookmarks=true,         
    unicode=false,          
    pdftoolbar=true,        
    pdfmenubar=true,        
    pdffitwindow=false,     
    pdfstartview={FitH},    
    pdftitle={My title},    
    pdfauthor={Author},     
    pdfsubject={Subject},   
    pdfcreator={Creator},   
    pdfproducer={Producer}, 
    pdfkeywords={keyword1, key2, key3}, 
    pdfnewwindow=true,      
    colorlinks=true,       
    linkcolor=blue,          
    citecolor=blue,
    filecolor=cyan,         
    urlcolor=magenta        
}

\SetCommentSty{mycommfont}

\SetKwInput{KwInput}{Input}                
\SetKwInput{KwOutput}{Output}              

\newcolumntype{M}[1]{>{\centering\arraybackslash}m{#1}}
\newcolumntype{P}[1]{>{\centering\arraybackslash}p{#1}}

\newcommand{\E}{\mathbb{E}}

\newcommand{\iid}{i.\@i.\@d.\ }

\theoremstyle{plain}
\newtheorem{lemma}{Lemma}
\newtheorem{proposition}{Proposition}

\newtheorem{theorem}{Theorem}
\newtheorem{remark}{Remark}

\newcommand\xqed[1]{%
  \leavevmode\unskip\penalty9999 \hbox{}\nobreak\hfill
  \quad\hbox{#1}}
\newcommand\demo{\xqed{$\blacksquare$}}

\begin{document}
\title{Scaling Law Tradeoff Between Throughput and Sensing Distance in Large ISAC Networks}

\author{Min Qiu,~\IEEEmembership{Member,~IEEE,} Ming-Chun Lee,~\IEEEmembership{Member,~IEEE,} Yu-Chih Huang,~\IEEEmembership{Senior Member,~IEEE,} and Jinhong Yuan,~\IEEEmembership{Fellow,~IEEE,}


\thanks{

Min Qiu and Jinhong Yuan are with the School of Electrical Engineering and Telecommunications, University of New South Wales, Sydney, NSW, 2052 Australia (e-mail: min.qiu@unsw.edu.au; j.yuan@unsw.edu.au).
Ming-Chun Lee and Yu-Chih Huang are with the Institute of Communications Engineering, National Yang Ming Chiao Tung University, Hsinchu 300, Taiwan (e-mail: mingchunlee@nycu.edu.tw; jerryhuang@nycu.edu.tw).
}
}

\maketitle

\begin{abstract}
In this paper, we investigate the fundamental tradeoff between communication and sensing performance of \emph{ad hoc} integrated sensing and communication (ISAC) wireless networks. Specifically, we consider that $n$ nodes are randomly located in an extended network with area $n$ and transmit ISAC signals. Under the pure path loss channel gain model and the condition that the transmission power scales according to the communication distance, we fully characterize the optimal scaling law tradeoff between throughput and sensing distance by proposing an achievable scheme and proving its converse. Our results can be interpreted as follows: by reducing the throughput by a factor of a function of $n$, the sensing range order improves according to the same function of $n$, raised to the power of the ratio between the path loss factors in communication and sensing. We prove that the same result also holds true for ISAC networks with random fading, despite the uncertainty on the connectivity and power level created by random fading. In addition, we show that the scaling law tradeoff cannot be improved by allowing the transmission power and communication distance to scale freely. To the best of our knowledge, this is the first work formally formulating and characterizing the communication and sensing performance scaling law tradeoff of \emph{ad hoc} ISAC networks.
\end{abstract}


\section{Introduction}
\IEEEPARstart{I}{Integrated} sensing and communication (ISAC) has emerged as a promising paradigm in 6G to enhance both communication and sensing capabilities \cite{10536135}. In ISAC, messages are conveyed to the receiver while sensing information is extracted from the scattered echoes at the same time. By employing unified waveforms and sharing the same wireless resources and infrastructures, ISAC can significantly reduce the energy and spectrum cost compared to separate communication and sensing systems on shared wireless resources \cite{9737357}.

The research on ISAC has attracted vibrant interest in both academia and industries. Numerous studies have explored key technologies in ISAC, such as waveform design \cite{9924202,9940978} and beamforming optimization \cite{9540344,10086626}. Meanwhile, several works have demonstrated the benefits of combining ISAC with other emerging technologies for 6G, e.g., orthogonal time-frequency space modulation \cite{9724198}, cell-free massive multiple-input multiple-output (MIMO) \cite{10000730}, rate-splitting multiple access \cite{10562043}, etc. Fundamentally, due to the conflicts in the purposes of communications and sensing, there exists a tradeoff between them in ISAC \cite{9705498}. Significant efforts have been made to characterize the fundamental tradeoff between sensing and communication performance in an information theoretic perspective from point-to-point channels to multi-terminal channels, e.g., see \cite{8437621,9785593,9787809,10147248}. That being said, most works focus on the design and analysis of ISAC at the link level. On the other hand, ISAC at the network level has not received much attention.

Compared to link-level (single-cell) ISAC, ISAC networks can expand the communication and sensing range, improve the quality of service, and obtain richer target information \cite{10726912,10735119,10769538}. However, characterizing the exact performance limit of large-size networks has been a long-standing open problem. A good starting point is to study how the performance scales with the network or node numbers, which can provide useful insights for system designers. As an example, \cite{10769538} introduced a cooperative ISAC network and studied the scaling of sensing accuracy Cramer-Rao bound with the number of ISAC base stations. Yet, the tradeoff between communication and sensing performance scaling laws has not been fully understood. In addition, such a cooperative protocol requires centralized control, which could be challenging to implement in large-scale deployment of ISAC transceivers. In contrast, \emph{ad hoc} wireless networks without any centralized control may be more favorable in these scenarios. The capacity scaling law of wireless networks has been extensively studied in the literature \cite{825799,1354532,4106120,4957662,6587994,LEI2024110696}, where the emphasis was on how the network performance can scale optimally with the number of nodes or the network size. Most notably, it was proved that for an \emph{ad hoc} wireless network with $n$ immobile transceiver nodes, simple multi-hop transmission, and treating interference as noise decoding, the optimal end-to-end communication throughput per node is in the order $1/\sqrt{n}$. This implies that as $n$ increases, the throughput per node will eventually diminish, while the overall system throughput will grow without bound. That said, the scaling law of the network performance changes significantly when incorporating other techniques, e.g., MIMO \cite{1638664,10.1145/1288107.1288124} and caching \cite{8094948,8809411}, in wireless networks. In this regard, it is expected that equipping every transceiver in a wireless network with sensing functionalities could also have a huge impact on the scaling law behavior.

Despite the above progress, the fundamental network analysis of ISAC networks with all transceivers performing ISAC tasks remains lacking. Although the current studies on ISAC networks, e.g., \cite{10735119,10769538}, have provided some guidelines in optimizing the network performance given specific scenarios and parameters, little is known about the fundamental properties and the ultimate performance gains of an ISAC network in general. This motivates us to study the communication and sensing performance scaling laws of large ISAC networks, which will help us better understand their capabilities and full potential. In this paper, we investigate the fundamental tradeoff between throughput and sensing distance scaling laws of \emph{ad hoc} ISAC networks, where $n$ nodes are randomly located in a network area of $n$ and transmit ISAC signals for performing communication and sensing tasks simultaneously. Moreover, we consider that all nodes employ point-to-point coding and \emph{naive} multi-hop transmission due to the ease of implementation and their successes in communication networks \cite{825799,1354532,4106120,4957662,6587994,LEI2024110696}. To the best of our knowledge, our work is the first attempt to formally formulate and rigorously characterize the tradeoff between communication and sensing performance scaling laws of ISAC networks. The main contributions are as follows.
\begin{itemize}
\item We first investigate the scaling law of ISAC networks under a pure path loss and absorption channel gain model with fixed path loss exponents. In this case, we show that it is necessary to scale the transmit power to improve the sensing distance scaling law. This, however, inevitably increases the interference level in the network. To address this issue, we introduce a natural condition that scales the power according to the communication distance to avoid unnecessary interference and power consumption in the network. Under this condition, we propose a new achievable scheme by constructing a highway system to delicately guarantee sufficient connectivity and bounded interference at each communication and sensing receiver, respectively, as the network size and node numbers go large. The resultant scaling law tradeoff can be interpreted as follows: by reducing the throughput by a factor of a function of $n$, the sensing distance order increases based on the same function of $n$, raised to the power of the ratio between the communication and sensing path loss exponents. In addition, we prove the optimality of the tradeoff by showing that the converse bound matches the achievability bound.
\item We explore an alternative approach by revoking the aforementioned condition and analyze the scaling law tradeoff. Specifically, we allow the transmit power, communication distances, and the number of transmission time slots to scale freely with $n$. We show that doing so cannot further improve the throughput nor the sensing distance scaling law of the network, even with transmit power scaling arbitrarily fast and achieving vanishing interference at each communication and sensing receiver. On the other hand, our results reveal that the alternative approach consumes much more power to achieve the same scaling law tradeoff as our original approach.
\item We extend the proposed design to the scenario where the channel gain consists of both path loss and random fading. In this case, the randomness in the power of received and interference signals introduces uncertainty in the connectivity, communication, and sensing quality of the ISAC network. To cope with this uncertainty, we introduce new routing approaches to construct a fading highway system. We show that by exploiting the spatial diversity of the highway, sufficient connectivity exists in the network to allow a constant data rate and a scalable sensing distance order to be achievable for each ISAC node with high probability. We then proceed by showing that the ISAC network with random fading is capable of achieving the same scaling law tradeoff as that without random fading.
\item Numerical results are provided to demonstrate the superiority of the proposed scheme over the benchmark scheme based on naive time-division multiplexing and the impacts of the path loss exponents on the scaling law tradeoff. Interestingly, our results reveal a peculiar behavior: a larger path loss exponent in the communication channel results in a better scaling law tradeoff.
\end{itemize}

\subsection{Notations}
Throughout the paper, we use the following order notations. We write $f(n) = O(g(n))$ if $\lim\sup_{n\rightarrow \infty}\frac{|f(n)|}{g(n)} <\infty$. We also write $f(n) = \Theta(g(n))$ if $f(n) = O(g(n))$ and $g(n) = O(f(n))$. We write $f(n) = o(g(n))$ if $\lim_{n \rightarrow \infty}\frac{f(n)}{g(n)} = 0$ provided that $g(n)$ is nonzero. We also write $f(n) = \omega(g(n))$ if $g(n) = o(f(n))$. The natural logarithm and logarithm base 2 are resented by $\log$ and $\log_2$, respectively.

\section{System Model}\label{sec:model}

We consider an extended network (i.e., constant node density with network area growing linearly with node numbers) in a square $S_n=[0,\sqrt{n}]\times [0,\sqrt{n}]$ with area $n$. There are $n$ nodes randomly distributed in $S_n$ according to a Poisson point process of unit intensity. We are interested in the events that occur inside $S_n$ with high probability (w.h.p.); that is, with probability tending to one as $n \rightarrow \infty$ \cite{4106120}. Let $X_k$ denote the location as well as the identity of node $k$, where $1 \leq k \leq n$. Let $\mathcal{T}(m)$ be the subset of nodes simultaneously transmitting at time instant $m$. Each node sends its ISAC signal to perform both communication and sensing tasks at the same time.

As a first attempt to the problem of the scaling law tradeoff in ISAC networks, we assume for simplicity that all nodes use point-to-point coding and decoding, have the same transmit power $P$, and are equipped with a single antenna. For ease of understanding, we first consider the pure path loss channel gain model and ignore the small-scale fading. In Sec. \ref{sec:fading}, we address the problem with random channel fading combining with path loss in the channel gain.



\subsection{Communication Model}\label{sec:comm_mod}
We pick uniformly at random a matching of source-destination pairs, such that each source node only corresponds to one destination node. The transmission from each source node to its destination node is performed in a multihop fashion, where the relaying node at each hop decodes the message and retransmits it to the next node. Let $X_k$ and $X_{R(k)}$ be a pair of transmitting and receiving nodes, respectively. For the communication channel, we assume that the signal power decays with distance $d$ as $1/d^{\alpha_{\text{c}}}$, where $\alpha_{\text{c}}$ is the path loss exponent for the communication model\footnote{We assume $d \geq 1$ such that the received signal should not be gaining energy due to path loss.}. For a pair of transmitting and receiving nodes $X_k$ and $X_{R(k)}$, we let $r_{k}$ be the transmission distance between them, i.e., $r_k \triangleq |X_k-X_{R(k)}|$. Moreover, let $d_{i,R(k)} \triangleq |X_i-X_{R(k)}|$ be the distance between the interfering node $X_i$ to $X_{R(k)}$. Each receiver simply treats the interference as Gaussian noise. Direct communication between $X_k$ and $X_{R(k)}$ over a channel of unit bandwidth can be established at a rate of
\begin{align}\label{eq:comm_SINR}
R_k = \log_2\left(1+\frac{\frac{P}{(r_{k})^{\alpha_{\text{c}}}}}{N_0+ \sum_{i\neq k, i\in \mathcal{T}(m)}\frac{P}{(d_{i,R(k)})^{\alpha_{\text{c}}}}}\right),
\end{align}
where $N_0$ denotes the Gaussian noise power and $\sum_{i\neq k, i\in \mathcal{T}(m)}\frac{P}{(d_{i,R(k)})^{\alpha_{\text{c}}}}$ is the sum of interfering ISAC signal power at the receiver node $X_{R(k)}$.

We are interested in the throughput that is achievable, concurrently, for all source-destination pairs
in the network, i.e., the \emph{per-node theoughput} $\lambda(n)$. A throughput of $\lambda(n)$ bits per second is said to be feasible if every node can send $\lambda(n)$ bits per second on average to its chosen destination node. That is, there is a $T<\infty$ such that every node is capable of transmitting $T\lambda(n)$ bits to its destination node in every time interval $[(i-1)T,iT]$. For any source-destination pair, let $D(n)$ be the required number of hops that guarantees the packet of any source node to reach its destination node. 

\subsection{Sensing Model}
In addition to performing communication tasks, each transmitter node employs monostatic radar, such that it acts as the sensing receiver and detects its surrounding targets through the reflection of its transmitted ISAC signals. We assume that the sensing channel is \emph{independent} of the communication channel\footnote{We assume that the self-interference at the sensing receiver can be mitigated by exploiting the knowledge of the transmitted signals, the corresponding channel, and by using advanced signal processing approaches and appropriate radio frequency circuit designs \cite{radar_book,10735119}. The clutter of sensing is assumed to be random, and thus can be absorbed in the noise term; some other processing techniques might also be used to mitigate the clutter \cite{radar_book}.}. In radar sensing, the signal attenuation suffers from the two-hop path loss effect \cite{10735119,10851815}. Hence, we use a different notation $\alpha_{\text{s}}$ to denote the path loss exponent for the sensing channel. As a result, the signal power decays with distance $d$ as $1/d^{\alpha_{\text{s}}}$. We consider that node $X_k$ is able to detect any targets within a distance of $d_k$ if the SINR of the received ISAC signal is above some constant threshold $\beta_{\text{s}}>0$. That is,
\begin{align}\label{eq:sensing_SINR}
\frac{\frac{P\sigma^2_0}{(d_{k})^{\alpha_{\text{s}}}}}{N_0+ \sum_{i\neq k, i\in \mathcal{T}(m)}\frac{P}{(d_{i,k})^{\alpha_{\text{c}}}}}\geq \beta_{\text{s}},
\end{align}
where $\sigma_0$ is the target radar cross section (RCS), $d_{i,k}$ is the Euclidean distance between sensing receiver node $X_k$ and interfering node $X_i$ and note that $d_{i,k}\neq d_{i,R(k)}$. Here, $d_k$ is known as the \emph{sensing distance} or sensing range of node $X_k$. In addition, we assume that $\sigma_0$ is a non-zero constant. For example, we can treat $\sigma_0$ as the minimum RCS among that of all the targets to be detected in the network as it does not affect the scaling law analysis.

In this work, we are interested in how the sensing distance scales with the network size or node numbers. To this end, we define $d(n)$ to be the sensing distance of the network, which is given by
\begin{align}\label{eq:dk_prop1}
d(n) =\sup\left\{d_{\text{s}}: d_k \geq d_{\text{s}} \text{ w.h.p.}\right\}.
\end{align}

\section{Scaling Law Tradeoff of ISAC Networks}\label{sec3}
In this section, we investigate the communication and sensing performance in \emph{ad hoc} ISAC networks.

First, by rearranging \eqref{eq:sensing_SINR}, we obtain the sensing distance of node $X_k$ as
\begin{align}\label{eq:dk1}
d_{k} \leq 
(\beta_{\text{s}})^{-\frac{1}{\alpha_{\text{s}}}}\left(\frac{P\sigma^2_0}{N_0+ \sum_{i\neq k, i\in \mathcal{T}(m)}\frac{P}{(d_{i,k})^{\alpha_{\text{c}}}}}\right)^{\frac{1}{\alpha_{\text{s}}}}.
\end{align}
It can be seen that in order to increase the sensing distance, the only card $X_k$ could play is to increase the power $P$. However, this inevitably increases its interference to other nodes.
Specifically, assume that the interference is bounded, which is the goal for many communication schemes for \emph{ad hoc} networks in the literature, e.g., \cite{825799,4106120}. Then, if one wants the sensing distance to scale with the network size or node numbers, it is \emph{necessary} for $P$ to scale with $n$. i.e., $P=f(n)$ for some increasing function of $n$. However, such a power scaling at every node would lead to excessive interference at each receiver. To tradeoff between the two conflicting goals, we tailor the achievable scheme in \cite{4106120} specifically to the ISAC networks, in which each transmission is forced to hop farther according to the transmission power dictated by the sensing requirement, while allowing fewer nodes to transmit simultaneously for ensuring bounded interference. The tradeoff between throughput and sensing distance scaling law is presented in Theorem \ref{th1}.

\begin{theorem}\label{th1}
For the \emph{ad hoc} ISAC network in Sec. \ref{sec:model} with $\alpha_{\text{c}}>2$ and $P=f(n) = O(n^{\frac{\alpha_{\text{c}}}{2}})$, the optimal tradeoff between throughput and sensing distance scaling law is given by
\begin{align}\label{eq:tradeoff}
(\lambda(n),d(n)) = \left(\Theta\left((f(n))^{-\frac{1}{\alpha_{\text{c}}}}n^{-\frac{1}{2}}\right),\Theta\left((f(n))^{\frac{1}{\alpha_{\text{s}}}}\right) \right).
\end{align}
\end{theorem}

\begin{IEEEproof}
See Sec. \ref{sec:ach1} and Sec. \ref{sec:converse} for the achievability and converse proofs, respectively.
\end{IEEEproof}

When $f(n)$ is fixed to be a constant, we recover the optimal communication throughput as $\Theta(\frac{1}{\sqrt{n}})$ bits/s. In this case, the sensing distance order of the network is fixed and does not scale with the network size or node numbers. On the other hand, the sensing-optimal point is reached when $f(n) = \Theta(n^{\frac{\alpha_{\text{c}}}{2}})$, whereas the throughput reduces to $\Theta(\frac{1}{n})$. This means that only one node is allowed to transmit at a time, which we refer to as the \emph{pure} time-division multiplexing (TDM). In addition, we restrict $P$ to be no larger than the order $n^{\frac{\alpha_{\text{c}}}{2}}$ to ensure bounded interference. The case of $P=\omega(n^{\frac{\alpha_{\text{c}}}{2}})$ will be covered in Sec. \ref{sec4}.

\begin{remark}
In \eqref{eq:tradeoff}, the mapping between the values of $\lambda(n)$ and $d(n)$ is one-to-one. In other words, given $\alpha_{\text{c}}$, $\alpha_{\text{s}}$, and $n$, the whole tradeoff curve is deterministic, governed by the choice of $f(n)$. Although we consider the asymptotic scaling law tradeoff where $P\rightarrow \infty$ as $n \rightarrow \infty$, the key message to deliver is how the power order growth rate impacts the scaling behavior of throughput and sensing distance. Note that scaling $P$ is also necessary in some communication scenarios, e.g., pure TDM (see Sec. \ref{Sec:TDM}) and the case in \cite[Th. 4.7]{NET-002}.
\demo
\end{remark}

\subsection{Achievability}\label{sec:ach1}


We only focus on the case where $f(n)$ scales slower than the order $n^{\frac{\alpha_{\text{c}}}{2}}$, i.e., $f(n) = o(n^{\frac{\alpha_{\text{c}}}{2}})$. The case $f(n) = \Theta(n^{\frac{\alpha_{\text{c}}}{2}})$ belongs to pure TDM, which will be discussed in Sec. \ref{Sec:TDM}.

\subsubsection{Setup}\label{sec:setup}
\begin{figure}[t!]
	\centering
\includegraphics[width=0.6\linewidth]{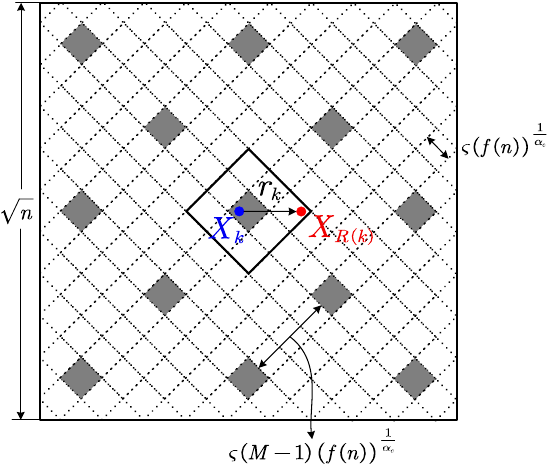}
\caption{The transmission pattern with $M=4$. Only the nodes in the grey subsquares are allowed to transmit simultaneously at a given time interval.}
\label{fig:grid}
\end{figure}

As shown in Fig. \ref{fig:grid}, we first partition square $S_n$ into $45^{\circ}$-angled subsquares $s_i$ of a side length $\varsigma (f(n))^{\frac{1}{\alpha_{\text{c}}}}$, where $\varsigma>0$ is a constant. The transmitter node $X_k$ selects the receiver node in one of its eight neighboring subsquares such that the communication distance satisfies $r_k \leq 2\sqrt{2}\varsigma (f(n))^{\frac{1}{\alpha_{\text{c}}}}$. Moreover, the side length of the subsquare within which the receiver node is located is $3\varsigma (f(n))^{\frac{1}{\alpha_{\text{c}}}}$. As we will show in Sec. \ref{sec:SINR}, this ensures that the distances between the interfering nodes and receiving nodes are large enough to achieve bounded interference at each receiver. Next, we use TDM to divide the total transmission time into $M^2$ slots, where $M>2$ is an integer constant. In other words, there are $M^2$ node groups that are not interfering with each other. Within each group, any neighboring subsquares with active nodes are separated at a distance at least $(M-1)$ subsquares away, i.e., the Euclidean distance between them is at least $\varsigma (M-1)(f(n))^{\frac{1}{\alpha_{\text{c}}}}$, as shown in Fig. \ref{fig:grid}. At time slot $m\in\{1,\ldots,M^2\}$, up to one node in each subsquare in subset $m$ transmits to a destination. Let $\mathcal{T}(m)$ be the subset of nodes simultaneously transmitting at time slot $m$. The proof of achievability consists of three steps.

\subsubsection{Bounding the Interference}\label{sec:SINR}
We need to ensure that the interference at each communication receiver node and sensing receiver node is bounded. Consider the transmitter node $X_k$ in Fig. \ref{fig:grid}. There are 8 interfering nodes in the first layer of interfering subsquares. These 8 interfering transmitting nodes are at least $\varsigma(M -2) (f(n))^{\frac{1}{\alpha_{\text{c}}}}$ away from communication receiver node $X_{R(k)}$. It is easy to see that there are $8l$ interfering nodes on the $l$-th layer with distance at least $\varsigma(lM -2) (f(n))^{\frac{1}{\alpha_{\text{c}}}}$ from the receiver node, for $l \geq 1$. One can see that $P$, $r_k^{\alpha_{\text{c}}}$, and $\min\{(d_{i,R(k)})^{\alpha_{\text{c}}}\}$ have the same order, i.e., $\Theta(f(n))$.

For any $k\in\mathcal{T}(m)$ and $m\in\{1,\ldots,M^2\}$, the interference at $X_{R(k)}$ can be upper bounded as
\begin{subequations}\label{eq:IN}
\begin{align}
\sum_{i\neq k, i\in \mathcal{T}(m)}\frac{P}{(d_{i,R(k)})^{\alpha_{\text{c}}}} \leq& \sum^{\infty}_{l=1}8l\frac{P}{\left(\varsigma(lM -2) (f(n))^{\frac{1}{\alpha_{\text{c}}}}\right)^{\alpha_{\text{c}}}} \\
=&\frac{8}{(\varsigma M)^{\alpha_{\text{c}}}}\sum^{\infty}_{l=1}\frac{l}{\left(l -\frac{2}{M} \right)^{\alpha_{\text{c}}}}.\label{eq:IN1}
\end{align}
\end{subequations}
By \cite[Remark 6.4]{NET-002}, the sum in the RHS of \eqref{eq:IN1} converges because $M>2$ and $\alpha_{\text{c}}>2$ are constants. Hence, the interference at any communication receiver at any transmission time slot is bounded. Similarly, we can follow the above steps to show that the interference at any sensing receiver at any time slot is also bounded, i.e.,
\begin{equation}\label{eq:IN_Xk}
\sum_{i\neq k, i\in \mathcal{T}(m)}\frac{P}{(d_{i,k})^{\alpha_{\text{c}}}}<\infty,\forall m\in\{1,\ldots,M^2\}.
\end{equation}

\subsubsection{Routing}\label{sec:routing}
We tailor the routing scheme in \cite{4106120} specifically to the considered ISAC networks. The routing consists of three phases, namely the draining phase, the highway phase, and the delivering phase. First, the packets are sent from each source node to the highway in \emph{one hop}. The highway consists of paths of hops that relay the packets at a constant rate either horizontally or vertically from one side of the network to the opposite side. The highway construction is based on a mapping to a bond percolation model \cite{Grimmett99}. In the last phase, packets are delivered from the highway to destination nodes in \emph{one hop}. In what follows, we show how the highway system is constructed. Note that in each phase, each transmitter node performs sensing tasks by using its transmitted or relaying ISAC signals.

\begin{figure}[t!]
	\centering
\includegraphics[width=0.5\linewidth]{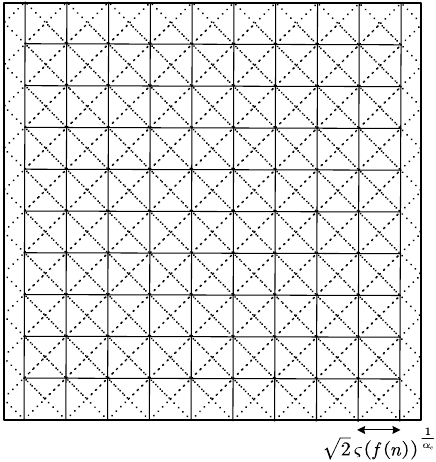}
\caption{Construction of the bond percolation model from the wireless network model in Fig \ref{fig:grid}.}
\label{fig:grid1}
\end{figure}

Let $|s_i|$ be the number of nodes inside subsquare $s_i$. We say the subsquare $s_i$ is \emph{open} if it contains at least one node, and \emph{closed} otherwise. Each subsquare in $S_n$ is open with probability
\begin{equation}\label{eq:p_open1}
\mathbb{P}_{\text{o}} \triangleq \mathbb{P}[|s_i| \geq 1]= 1-e^{-\varsigma^2 (f(n))^{\frac{2}{\alpha_{\text{c}}}}},
\end{equation}
independently of each other. We then map this network model to a bond percolation model on a square lattice \cite{Grimmett99}. For all the subsquares in $S_n$, mark half of them as horizontal subsquares and the other half as vertical subsquares. In each horizontal (resp. vertical) subsquare, we draw an edge across it horizontally (resp. vertically). Each edge is \emph{open} with probability $\mathbb{P}_{\text{o}}$ if its underlying subsquare is open, and \emph{closed} with probability $1-\mathbb{P}_{\text{o}}$ otherwise. In this way, we obtain a grid $B_n$ of horizontal and vertical edges as shown in Fig \ref{fig:grid1}. Each path on $B_n$ is formed by connected edges, and is \emph{open} if it contains only open edges.

\emph{3a) Highway Construction:} We start constructing the horizontal highway system. Define $\xi \triangleq \frac{\sqrt{n}}{\sqrt{2}\varsigma (f(n))^{\frac{1}{\alpha_{\text{c}}}}}$. We partition the network into horizontal rectangles $R_{\text{h}}$ of size $\sqrt{n} \times \frac{\kappa\sqrt{n}}{\xi}\log \xi$, where $\kappa>0$ is some constant. Each $R_{\text{h}}$ has $\xi \times \kappa\log \xi$ subsquares $s'_i$ of side length $\sqrt{2}\varsigma (f(n))^{\frac{1}{\alpha_{\text{c}}}}$, where $s'_i$ is formed by the edges of $s_i$. The following proposition shows that w.h.p. there are a sufficient number of disjoint open paths in each $R_{\text{h}}$ even when the transmission has to hop farther based on the transmission power dictated by the sensing requirement.

\begin{proposition}\label{th2}
For any constants $\kappa>0$ and $\varsigma>0$, there exists $\eta>0$ such that w.h.p. there are $\eta\kappa \log \frac{\sqrt{n}}{\sqrt{2}\varsigma (f(n))^{\frac{1}{\alpha_{\text{c}}}}}$ disjoint open paths inside each horizontal retangle $R_{\text{h}}$ of $S_n$.
\end{proposition}
\begin{IEEEproof}
The proof is in Appendix \ref{sec:proof_th2}.
\end{IEEEproof}

By Proposition \ref{th2}, we know that the total number of disjoint horizontal open paths in the network is $\eta\frac{\sqrt{n}}{\sqrt{2}\varsigma (f(n))^{\frac{1}{\alpha_{\text{c}}}}}$ w.h.p. Using a similar construction as above, we can also obtain that the total number of disjoint vertical open paths is $\eta\frac{\sqrt{n}}{\sqrt{2}\varsigma (f(n))^{\frac{1}{\alpha_{\text{c}}}}}$ w.h.p. In other words, the total number of open paths in the network is of the same order as that of the total number of subsquares $s'_i$ in $S_n$. Finally, all the horizontal and vertical open paths form the highway system. As we will see, the existence of such a highway system is beneficial to both communication and sensing, in the sense that a constant data rate and scalable sensing distance is achievable at every ISAC node w.h.p.


%
%


\subsubsection{Achievable Throughput and Sensing Distance}\label{sec:bound_acc}
In this section, we bound the throughput and sensing distance for each routing phase.

\emph{4a) Draining Phase:} In this phase, each source node transmits its signal to an entry point node in one of the open paths in the highway while performing its sensing task. The source node and the entry point node are located in the same horizontal rectangle $R_{\text{h}}$.

To analyze the communication rate between nodes, we follow the setup in Sec. \ref{sec:setup} and modify $r_k$ and $M$ as follows. From Sec. \ref{sec:routing}, we know that the vertical distance between the source node and the highway entry point is at most the width of $R_{\text{h}}$, which is $\frac{\kappa\sqrt{n}}{\xi}\log \xi$. This means that the communication distance between the source and the entry point $r_k$ is within $\frac{\sqrt{2}\kappa\sqrt{n}}{\xi}\log \xi$. Since the desired signal power reduces due to the increase in the transmission distance combined with path loss, we set number of the transmission time slots to be $M^2=4(\sqrt{2}\kappa\log\xi+1)^2$ to further reduce the number of simultaneous transmission nodes and the interference power. Following \eqref{eq:comm_SINR} and \eqref{eq:IN}, we derive the achievable communication rate as
\begin{subequations}\label{eq:draining_Rk}
\begin{align}
R_k = &\log_2\left(1+\frac{\frac{1}{\left(2\kappa\varsigma \log \xi\right)^{\alpha_{\text{c}}}}}{N_0+\frac{8}{\varsigma^{\alpha_{\text{c}}}\left(\sqrt{2}\kappa\log\xi+1\right)^{\alpha_{\text{c}}}}\sum^{\infty}_{l=1}\frac{l}{\left(2l-1\right)^{\alpha_{\text{c}}}}}\right) \\
=&\Theta\left(\left(\log\frac{\sqrt{n}}{\sqrt{2}\varsigma (f(n))^{\frac{1}{\alpha_{\text{c}}}}}\right)^{-\alpha_{\text{c}}}\right), \label{eq:draining_Rk1}
\end{align}
\end{subequations}
where \eqref{eq:draining_Rk1} follows because the sum $\sum^{\infty}_{l=1}\frac{l}{(2l-1)^{\alpha_{\text{c}}}}$ converges.
Since there are possibly more than one node in one subsquares, they have to take turns transmitting their signals to the entry point. Consequently, the per-node throughput in the draining phase is given by
\begin{subequations}\label{eq:throughput_drain}
\begin{align}
\lambda(n) =& \frac{R_k}{M^2\max\{|s_i|\}} \\
=&\frac{\Theta\left(\left(\log\frac{\sqrt{n}}{\sqrt{2}\varsigma (f(n))^{\frac{1}{\alpha_{\text{c}}}}}\right)^{-\alpha_{\text{c}}}\right)}{4(\kappa\log\xi+1)^2\cdot\varsigma f(n)^{\frac{1}{\alpha_{\text{c}}}}\sqrt{n} \left(\delta\log \frac{\sqrt{n}}{\sqrt{2}\varsigma f(n)^{\frac{1}{\alpha_{\text{c}}}}}\right)^{-\alpha_{\text{c}}-2}} \label{eq:drain_t1} \\
=& \Theta\left( \frac{1}{f(n)^{\frac{1}{\alpha_{\text{c}}}}\sqrt{n}}\right),\label{eq:drain_t2}
\end{align}
\end{subequations}
where \eqref{eq:drain_t1} follows by using \eqref{eq:draining_Rk} as well as Lemma \ref{lem:AN} in Appendix \ref{sec:lemma} to bound the number of nodes in $s_i$ with $\delta>0$ being some constant.

To compute the sensing distance order in the draining phase, we apply a similarly approach as in \eqref{eq:draining_Rk} to bound the interference at the sensing receiver. It is not difficult to see that the interference at $X_k$ is of the order $\Theta\left(\left(\log \frac{\sqrt{n}}{f(n)^{1/\alpha_{\text{c}}}}\right)^{-\alpha_{\text{c}}}\right)$ for all $k \in \mathcal{T}(m)$ in the draining phase, which vanishes as $n \rightarrow \infty$ due to the reduced number of simultaneous transmission nodes. By \eqref{eq:dk_prop1} and \eqref{eq:dk1} while letting the sensing SINR equals $\beta_{\text{s}}$, w.h.p. the sensing distance in the draining phase is
\begin{subequations}\label{eq:dk_drain}
\begin{align}
d(n) =& (\beta_{\text{s}})^{-\frac{1}{\alpha_{\text{s}}}}\frac{(f(n))^{\frac{1}{\alpha_{\text{s}}}}}{\left(N_0+ \Theta\left(\left(\log \frac{\sqrt{n}}{f(n)^{\frac{1}{\alpha_{\text{c}}}}}\right)^{-\alpha_{\text{c}}}\right)\right)^{\frac{1}{\alpha_{\text{s}}}}}  \\
= & \Theta\left( (f(n))^{\frac{1}{\alpha_{\text{s}}}} \right).\label{eq:dk_no_I}
\end{align}
\end{subequations}

\emph{4b) Highway Phase:} In this phase, the ISAC signals are first transmitted with multihop relaying along the horizontal highway path from the entry point to the crossing point with the vertical highway path associated with the destination node. Then, those signals are still transmitted with multihop relaying from the crossing point to the exit point of the vertical highway.

We first apply the setup in Sec. \ref{sec:setup} to analyze the communication rate. Since the interference at every communication receiver in this phase is bounded according to \eqref{eq:IN}, the communication rate between hops is given by
\begin{equation}\label{eq:Rk_highway}
R_k = \log_2 \left(1+\frac{\frac{1}{(2\sqrt{2}\varsigma)^{\alpha_{\text{c}}}}}{N_0+ \frac{8}{(\varsigma M)^{\alpha_{\text{c}}}}\sum^{\infty}_{l=1}\frac{l}{\left(l -\frac{2}{M} \right)^{\alpha_{\text{c}}}}} \right),
\end{equation}
which can be lower bounded by a constant.

Clearly, the communication rate in the highway phase is higher than that in the draining phase in \eqref{eq:draining_Rk}. However, the throughput needs to take into account the time penalty due to multihop transmission, whereas the tranission in the draining phase takes one hop only. By \cite[Lemma 2]{4106120}, there are at most $2w\sqrt{n}$ nodes in a slice of width $w$. Since only one node each subsquare $s_i$ of side length $\varsigma (f(n))^{\frac{1}{\alpha_{\text{c}}}}$ is allowed to transmit at one time interval, the number of hops satisfies $D(n) = \Theta\left(\frac{\sqrt{n}}{(f(n))^{\frac{1}{\alpha_{\text{c}}}}}\right)$. Moreover, given that the number of subsquares can simultaneously transmit is $\frac{n}{\varsigma^2 (f(n))^{\frac{2}{\alpha_{\text{c}}}}M^2}$, each highway node transmits at least once in every $\varsigma^2 (f(n))^{\frac{2}{\alpha_{\text{c}}}}M^2$ time slots. Hence, w.h.p. the per-node throughput is
\begin{subequations}
\begin{align}\label{lambdan}
\lambda(n) =& \frac{R_k}{\varsigma^2(f(n))^{\frac{2}{\alpha_{\text{c}}}}M^2D(n)}  \\
\overset{\eqref{eq:Rk_highway}}{=}&\Theta\left( \frac{1}{(f(n))^{\frac{1}{\alpha_{\text{c}}}}\sqrt{n}}\right).\label{eq:13b}
\end{align}
\end{subequations}

For the sensing distance, we note from Sec. \ref{sec:SINR} that the interference at every sensing receiver at any time slot of the highway phase is bounded. Using \eqref{eq:dk_prop1} and \eqref{eq:dk1} again, one can show that w.h.p. the sensing distance order with bounded interference is the same as that with vanishing interference
in \eqref{eq:dk_no_I}
\begin{align}\label{eq:dk2}
d(n) 
=  \Theta\left( (f(n))^{\frac{1}{\alpha_{\text{s}}}} \right).
\end{align}

\emph{4c) Delivering Phase:} In this phase, ISAC signals are transmitted from the exit point node of the vertical highway to the destination node. Since the delivery phase is symmetric to the draining phase, one can follow Sec. \ref{sec:bound_acc}a to analyze the throughput and the sensing distance. The only difference is that the direction of signal transmission is vertical. As a result, the throughput and sensing distance as the same as that in \eqref{eq:throughput_drain} and \eqref{eq:dk_drain}, respectively.

We note that the throughput and sensing distance scaling laws are the same for all three routing phases. Therefore, the results in Theorem \ref{th1} follow.

\begin{remark}\label{remark1}
For sensing, there exists a tradeoff between sensing distance and sensing frequency, where the sensing frequency is defined as the average number of times that a node performs sensing over the total time slots. In the proposed scheme, the sensing frequency is in the order $O\left(\frac{1}{\varsigma^2(f(n))^{\frac{1}{\alpha_{\text{c}}}}M^2}\right)$. To allow the sensing distance to increase, nodes should perform sensing less frequently to reduce the interference. However, finding the optimal tradeoff between sensing distance and sensing frequency is non-trivial as it involves scheduling design. This will be investigated in our future work.
\demo
\end{remark}

\subsection{Converse}\label{sec:converse}
The converse proof is built upon the results established for \emph{ad hoc} wireless networks with independently and uniformly distributed nodes under the \emph{Protocol Model} \cite{NET-002}\footnote{Note that the number of nodes follows a binomial distribution, which becomes Poisson distribution as $n \rightarrow \infty$. Since we focus on high probability events, the converse proof for the case of the Binomial point process also holds for the Poisson point process as long as $n \rightarrow \infty$ at the same order.}. Before we proceed, we briefly describe the model.

In the Protocol Model, for any $i\neq k$ and $i,k\in \mathcal{T}$, node $X_k$ can successfully transmit $W$ bits/s to node $X_{R(k)}$ if $|X_i-X_{R(k)}| \geq (1+\Delta)|X_k-X_{R(k)}|$ for every other node $X_i$ simultaneously transmit at the same time, where $\Delta>0$ is a constant quantifying a guard zone for $X_{R(k)}$ to ensure no destructive interference from the neighboring node $X_i$. In addition, we define the maximum transmission range as $r_n$, such that $r_k \leq r_n,\forall k \in \mathcal{T}$. The converse proof consists of the following three steps.

\subsubsection{Maximum Number of Simultaneous Transmissions}\label{sec:max_sim_trans}
We start with an important result for the Protocol Model. By \cite[Lemma 5.14]{NET-002}, under the Protocol Model with $n$ nodes located in a square with area $n$, the maximum number of simultaneous transmissions is no more than
\begin{align}
\frac{16n}{\Delta^2r^2_n}. \label{eq:max_trans}
\end{align}

Then, we need to apply the above result to our network model. We note that node $X_k$ can successfully transmit a constant rate to node $X_{R(k)}$ if there exits a constant threshold $\beta_{\text{c}}>0$ such that the signal-to-interference ratio (SIR) satisfies $\frac{P}{(r_k)^{\alpha_{\text{c}}}}/(\frac{P}{(d_{i,R(k)})^{\alpha_{\text{c}}}}) \geq \beta_{\text{c}}$ for every other node $X_i$ simultaneously transmit at the same time. This can be achieved if the interference is bounded, see Sec. \ref{sec:SINR}. Rearranging the above condition for the SIR, we get $|X_i-X_{R(k)}| \geq (\beta_{\text{c}})^{\frac{1}{\alpha_{\text{c}}}}|X_k-X_{R(k)}|$, which gives $\Delta = (\beta_{\text{c}})^{\frac{1}{\alpha_{\text{c}}}}-1$. Thus, the number of simultaneous transmissions feasible for the Protocol Model in \eqref{eq:max_trans} is also feasible for the considered network model in Sec. \ref{sec:comm_mod}.


\subsubsection{Transmission Range Requirement}
In the communication model, certain constraints are required on the transmission range for the feasibility of any throughput in the random network under the Protocol Model. By \cite[Corollary 5.1]{825799}, the necessary condition for any feasible throughput in the extended network under the Protocol Model is the \emph{absence} of isolated nodes, which requires the transmission range satisfies
\begin{align}
r_n \geq c_1\sqrt{\log n},
\end{align}
for some constant $c_1>0$.

As we have seen in Sec. \ref{sec:max_sim_trans}, the Protocol Model is a rather restricted model compared to ours. Hence, the above lower bound on the transmission range is also applicable to the channel model under consideration. In what follows, we use the transmission range constraint to obtain the upper bounds for throughput and sensing distance.

\subsubsection{Throughput and Sensing Distance Upper Bound}
We first bound the throughput. Let $\bar{L}$ be the mean distance between a pair of uniformly and independently chosen source and destination nodes within $S_n$. Then, the number of hops is $\frac{\bar{L}}{r_n} = \Theta(\frac{\sqrt{n}}{r_n})$. Since each node generates packets at rate $\lambda(n)$, this means that the bits per second being transmitted by the whole network is at least $n\lambda(n)\frac{\bar{L}}{r_n}$. To ensure that the network can support such data traffic, it is required that
\begin{equation}
n\lambda(n)\frac{\bar{L}}{r_n} \overset{\eqref{eq:max_trans}}{\leq} W \frac{16}{\Delta^2r^2_n},
\end{equation}
which gives
\begin{equation}
\lambda(n)\leq \frac{16W}{\left((\beta_{\text{c}})^{\frac{1}{\alpha_{\text{c}}}}-1\right)^2r_n\bar{L}}=  \Theta\left(\frac{1}{r_n\sqrt{n}}\right).\label{eq:lambda}
\end{equation}

We now bound the sensing distance by using the constraint on the communication range. In general, the communication range of a node is not directly related to its sensing distance. However, it is important to realize that one of the necessary conditions to guarantee a constant communication rate is to avoid the received signal power $\frac{P}{(r_n)^{\alpha_{\text{c}}}}$ vanishes as $n \rightarrow \infty$, apart from the bounded interference condition. In this regard, the transmit power should satisfy $P = \Theta((r_n)^{\alpha_{\text{c}}})$. Hence, we can rearrange \eqref{eq:dk1} as follows
\begin{subequations}
\begin{align}
d(n) \leq & 
(r_n)^{\frac{\alpha_{\text{c}}}{\alpha_{\text{s}}}}(\beta_{\text{s}})^{-\frac{1}{\alpha_{\text{s}}}}\left(\frac{\frac{P}{(r_n)^{\alpha_{\text{c}}}}}{N_0+ \sum_{i\neq k, i\in \mathcal{T}}\frac{P}{(d_{i,k})^{\alpha_{\text{c}}}}}\right)^{\frac{1}{\alpha_{\text{s}}}} \label{eq:dk_upper0} \\
= & \Theta\left((r_{n})^{\frac{\alpha_{\text{c}}}{\alpha_{\text{s}}}}\right), \label{eq:dk_upper}
\end{align}
\end{subequations}
where \eqref{eq:dk_upper} follows since $\frac{P}{(r_n)^{\alpha_{\text{c}}}}$ and the interference at each sensing receiver is bounded (see \eqref{eq:IN_Xk}).



Combining \eqref{eq:lambda} and \eqref{eq:dk_upper}, we obtain the tradeoff between throughput and sensing distance scaling law as
\begin{align}\label{eq:tradeoff_converse}
(\lambda(n),d(n))=\left(\Theta\left(\frac{1}{r_n\sqrt{n}}\right),\Theta\left((r_{n})^{\frac{\alpha_{\text{c}}}{\alpha_{\text{s}}}}\right) \right),
\end{align}
where we note that the transmission distance satisfies $c_1 \sqrt{\log n} \leq r_n \leq c_2\sqrt{n}$ for some constants $c_1>0$ and $c_2>0$. We can also optimistically assume that $r_n \geq c_3$ for some constant $c_3>0$ to obtain a throughput upper bound up to $\lambda(n) = \Theta(\frac{1}{\sqrt{n}})$. By letting $r_n = (f(n))^{\frac{1}{\alpha_{\text{c}}}}$, the tradeoff in \eqref{eq:tradeoff_converse} becomes \eqref{eq:tradeoff}, where the order of $f(n)$ is no larger than the order $n^{\frac{\alpha_{\text{c}}}{2}}$ due to the constraint $r_n \leq c_2\sqrt{n}$.

\section{An Alternative Scaling Law Tradeoff}\label{sec4}
In Sec. \ref{sec3}, we focused on achieving bounded interference at each communication and sensing receiver by restricting $P$ to scale no faster than $n^{\frac{\alpha_{\text{c}}}{2}}$. In this section, we relax this constraint to explore the possibility of achieving a better scaling law tradeoff in ISAC networks.

\subsection{The Pure TDM Scheme}\label{Sec:TDM}
In this section, we investigate the scaling laws of throughput and sensing distance of ISAC networks operating under the pure TDM scheme.

First, let the transmit power be $P = f(n)$. For this scheme, only one node is allowed to transmit at any given time slot such that each receiver node is interference-free. Each source node directly communicates to the destination node without using multihop, at the cost of huge path loss. In particular, the average distance of a randomly picked source-destination pair in $S_n$ is in the order of $\sqrt{n}$. The throughput is given by
\begin{align}
\lambda(n)=\left\{ \begin{array}{ll}
\Theta(f(n)n^{-1-\frac{\alpha_{\text{c}}}{2}})& \text{if }f(n) = o(n^{\frac{\alpha_{\text{c}}}{2}})\\
\Theta(n^{-1})& \text{if }f(n) = \Theta(n^{\frac{\alpha_{\text{c}}}{2}})\\
\Theta(n^{-1}\log f(n))& \text{if }f(n) = \omega(n^{\frac{\alpha_{\text{c}}}{2}})\\
\end{array} \right..
\end{align}

In addition, pure TDM always achieves a sensing distance of $d(n) = (f(n))^{\frac{1}{\alpha_{\text{s}}}}$ due to the absence of interference.

Compared to \eqref{eq:tradeoff}, pure TDM is strictly suboptimal in terms of degraded throughput when $P=O(n^{\frac{\alpha}{2}})$. In fact, it requires that the power should at least scale exponentially with $n$, i.e., $f(n) = \omega(e^{\sqrt{n}})$, to beat the $\Theta(\frac{1}{\sqrt{n}})$ throughput achieved by constant power. Although pure TDM achieves the same sensing distance scaling law in \eqref{eq:tradeoff}, the sensing frequency drops to the lowest, i.e., $\frac{1}{n}$. Essentially, the performance of the pure TDM scheme is more about a single node's performance, which does not really demonstrate what an ISAC network is capable of.

%

\subsection{An Alternative Achievable Scheme}\label{sec:alternative_scheme}
Now we turn to another achievable scheme instead of using pure TDM. In Sec. \ref{sec3}, we introduced the function $f(n)$ to control the asymptotic behavior of the transmit power, communication distance, and partitioned subsquare side length. A natural question arises: can we achieve a better tradeoff by relaxing the constraints on the scalings of those quantities? In this section, we show a negative result, indicating that a better tradeoff cannot be achieved by allowing $P$, $r_k$, and $d_{i,R(k)}$ to scale differently with $n$, at least with the techniques presented in this paper.

\subsubsection{Setup}\label{sec4_setup}
For simplicity, we use the same routing protocol as in Sec. \ref{sec:ach1} and focus on the performance in the highway phase only. We adopt the setup in Sec. \ref{sec:setup} with the following modification. Let $f_1(n)$, $f_2(n)$, and $f_3(n)$ be three increasing functions of $n$. Let $P = f_1(n)$ which can scale faster than the communication distance. Next, we divide $S_n$ into subsquares of a side length $\varsigma f_2(n)$, where $c_4 \leq f_2(n) < c_5\sqrt{n}$ and $c_4>0$ and $c_5>0$ are some constants. In this case, node $X_k$ choose its receiver node $X_{R(k)}$ in one of its eight neighboring subsquares such that the communication distance between nodes satisfies $r_k \leq 2\sqrt{2}\varsigma f_2(n)$. Moreover, we consider $f_2(n) = o((f_1(n))^{\frac{1}{\alpha_{\text{c}}}})$ to allow the received signal power $\frac{P}{(r_k)^{\alpha_{\text{c}}}}$ to scale with $n$ rather than being bounded by a constant as in Sec. \ref{sec:ach1}. The total transmission time is divided into $M^2=(f_3(n))^2$ slots, where $2 < f_3(n) < \frac{\sqrt{n}}{\sqrt{2}\varsigma f_2(n)}+2$. The upper bound of $f_3(n)$ is due to the distance constraint $d_{i,{R(k)}}<\sqrt{n}$. In addition, we only consider $f_2(n)=o(\sqrt{n})$ and $f_3(n)=o(\sqrt{n})$; otherwise the case becomes the pure TDM scheme in Sec. \ref{Sec:TDM}.

Based on the proposed setup and following \eqref{eq:IN}, we can derive the interference at each communication receiver $X_{R(k)}$ at any time slot of the highway phase as
\begin{align}\label{eq:IN_n}
\sum_{i\neq k, i\in \mathcal{T}(m)}\frac{P}{(d_{i,R(k)})^{\alpha_{\text{c}}}} \leq \frac{8f_1(n)}{(\varsigma f_2(n) f_3(n))^{\alpha_{\text{c}}} }\sum^{\infty}_{l=1}\frac{l}{\left(l -\frac{2}{f_3(n)}\right)^{\alpha_{\text{c}}}},
\end{align}
where the sum $\sum^{\infty}_{l=1}\frac{l}{\left(l -\frac{2}{f_3(n)}\right)^{\alpha_{\text{c}}}}$ in the RHS of \eqref{eq:IN_n} converges according to \cite[Remark 6.4]{NET-002} since $\alpha_{\text{c}}>2$ and $f_3(n)>2$. Similarly, one can follow the above steps to show that the interference at each sensing receiver $X_k$ at any time satisfies
\begin{align}\label{eq:IN_sensing1}
\sum_{i\neq k, i\in \mathcal{T}(m)}\frac{P}{(d_{i,k})^{\alpha_{\text{c}}}} \leq \frac{8f_1(n)}{(\varsigma f_2(n) f_3(n))^{\alpha_{\text{c}}} }\sum^{\infty}_{l=1}\frac{l}{\left(l -\frac{1}{f_3(n)}\right)^{\alpha_{\text{c}}}}.
\end{align}

Depending on the orders of the functions, the interference at each communication receiver and sensing receiver can either scale with $n$ or vanish as $n \rightarrow \infty$.

\subsubsection{Throughput Scaling Law}
We first compute the communication rate between nodes. Substituting \eqref{eq:IN_n} into \eqref{eq:comm_SINR}, we get
\begin{subequations}
\begin{align}
R_k
=& \log_2\left(1+\frac{\frac{f_1(n)}{(f_2(n))^{\alpha_{\text{c}}}}}{N_0+ \frac{8f_1(n)}{(\varsigma f_2(n) f_3(n))^{\alpha_{\text{c}}} }\sum^{\infty}_{l=1}\frac{l}{\left(l -\frac{2}{f_3(n)}\right)^{\alpha_{\text{c}}}}}\right) \\
  =& \Theta\left(\frac{f_1(n)(f_3(n))^{\alpha_{\text{c}}}}{(f_2(n)f_3(n))^{\alpha_{\text{c}}}+f_1(n)}\right). \label{eq:Rk_diff1}
\end{align}
\end{subequations}

To proceed further, we consider two conditions, namely C1: $f_2(n)f_3(n) = o((f_1(n))^{\frac{1}{\alpha_{\text{c}}}})$ and C2: $f_1(n) = O((f_2(n)f_3(n))^{\alpha_{\text{c}}})$. Note that C1 includes the case that the transmit power order can scale arbitrarily fast with $n$, while C2 includes the case of vanishing interference as $n \rightarrow \infty$. It is worth noting that the order of the communication rate in \eqref{eq:Rk_diff1} scales with $n$ rather than being a constant, with either C1 or C2 holds\footnote{Since the data rate in this case scales with $n$ rather than being lower bounded by a constant, one should not follow Sec. \ref{sec:converse} to derive the converse in this case.}.

%

Having derived the communication rate, we then follow the steps in Sec. \ref{sec:bound_acc} to obtain the throughput. Given that the number of hops satisfies $D(n) = \Theta\left(\frac{\sqrt{n}}{f_2(n)}\right)$ and the number of subsquares can simultaneously transmit is $\frac{n}{(\varsigma f_2(n)f_3(n))^2}$, w.h.p. the throughput is
\begin{subequations}\label{eq:thr_2}
\begin{align}
\lambda(n) =& \frac{R_k}{(\varsigma f_2(n)f_3(n))^2D(n)}  \\
=& \Theta\left(\frac{ \log\frac{f_1(n)(f_3(n))^{\alpha_{\text{c}}}}{(f_2(n)f_3(n))^{\alpha_{\text{c}}}+f_1(n)}}{f_2(n)( f_3(n))^2\sqrt{n}}\right)\\
=&\left\{ \begin{array}{ll}
\Theta\left(\frac{ \log f_3(n)}{f_2(n)( f_3(n))^2\sqrt{n}}\right)& \text{if C1 holds}\\
\Theta\left(\frac{ \log\frac{f_1(n)}{(f_2(n))^{\alpha_{\text{c}}}}}{f_2(n)( f_3(n))^2\sqrt{n}}\right)& \text{if C2 holds}\\
\end{array} \right. \label{eq:array1}\\
=&o\left(\frac{1}{f_2(n)f_3(n)\sqrt{n}}\right), \label{eq:thr_3}
\end{align}
\end{subequations}
where \eqref{eq:thr_3} follows that $\log f_3(n)<f_3(n)$ if C1 holds and $\log \frac{f_1(n)}{(f_2(n))^{\alpha_{\text{c}}}}=o( f_3(n))$ if C2 holds.

Observe that the throughput scaling law in \eqref{eq:thr_3} is worse than that in \eqref{eq:tradeoff}. In particular, if $P$ scales arbitrarily fast with $n$ as in C1, the interference also scales with $n$. In this case, more time slots are required to ensure that the received signal power $\frac{P}{(r_k)^{\alpha_{\text{c}}}}$ should scale faster than the interference with $n$. Hence, the throughput scaling law suffers from a time penalty introduced by the increasing transmission time slot. On the other hand, if we aim for vanishing interference when C2 holds, then it is necessary to scale the number of transmission time slots with $n$. This also leads to reduced throughput.

%


\subsubsection{Sensing Distance Scaling Law}
Since the interference at each sensing receiver at any time slot of the highway phase can be upper bounded by the same constant, we thus obtain the sensing distance by substituting \eqref{eq:IN_sensing1} into \eqref{eq:dk1} as
\begin{subequations}
\begin{align}
d(n) = &
(\beta_{\text{s}})^{-\frac{1}{\alpha_{\text{s}}}}\left(\frac{f_1(n)}{N_0+\frac{8f_1(n)}{(\varsigma f_2(n) f_3(n))^{\alpha_{\text{c}}} }\sum^{\infty}_{l=1}\frac{l}{\left(l -\frac{1}{f_3(n)}\right)^{\alpha_{\text{c}}}}}\right)^{\frac{1}{\alpha_{\text{s}}}} \\
=&\Theta\left(\left(\frac{f_1(n)(f_2(n)f_3(n))^{\alpha_{\text{c}}}}{(f_2(n)f_3(n))^{\alpha_{\text{c}}}+f_1(n)}\right)^{\frac{1}{\alpha_{\text{s}}}}\right) \\
=&\left\{ \begin{array}{ll}
\Theta\left((f_2(n)f_3(n))^{\frac{\alpha_{\text{c}}}{\alpha_{\text{s}}}}) \right) &\text{if C1 holds}\\
\Theta\left((f_1(n))^{\frac{1}{\alpha_{\text{s}}}}) \right)& \text{if C2 holds}\\
\end{array} \right.. \label{eq:ds_2}
\end{align}
\end{subequations}

It can be seen that the sensing distance scaling law in \eqref{eq:ds_2} is not better than that in \eqref{eq:dk2}. If $P$ scales arbitrarily fast with $n$ under condition C1, the interference at the sensing receiver becomes unbounded. Consequently, the advantage of using excessive transmit power is lost completely. Thus, the sensing distance order in \eqref{eq:ds_2} becomes irrelevant to the order of $P$ under condition C1. In addition, for C2 achieving vanishing interference at each sensing receiver requires that $P$ scales slower than $d_{i,R(k)}$ and $d_{i,k}$ with $n$. As a result, the sensing distance scaling law is essentially constrained by the node distance scaling, which cannot be larger than the order $(\sqrt{n})^{\frac{\alpha_{\text{c}}}{\alpha_{\text{s}}}}$.

In summary, the tradeoff obtained in this section by scaling the transmit power, the node distance, and the number of transmission time slots differently is worse than that in \eqref{eq:tradeoff} due to the degraded throughput in \eqref{eq:thr_3}. In other words, achieving vanishing interference or scaling the communication rate with $n$ does not improve the throughput compared to the case in \eqref{eq:thr_3}. In addition, under condition C1 the power consumption of each node is much larger than that of the case in presented in Sec. \ref{sec3}. On the other hand, under condition C2, using the same power scaling as in Sec. \ref{sec3} achieves the same sensing distance order but worse throughput.

\begin{remark}
The scaling law tradeoff cannot be further improved even if we fix $f_2(n)$ or $f_3(n)$ or both to be constants. First, it is not difficult to see that the scaling laws \eqref{eq:thr_3} and \eqref{eq:ds_2} still hold when $f_2(n)$ is a constant while $f_3(n)$ remains to be an increasing function of $n$. If $f_3(n)$ is a constant while $f_2(n)$ is not, one obtains the same scaling law tradeoff as in \eqref{eq:tradeoff} at the cost of much higher power consumption because we have $f_2(n) = o((f_1(n))^{\frac{1}{\alpha_{\text{c}}}})$ in the setup. Finally, when both $f_2(n)$ and $f_3(n)$ are constants, the ISAC network achieves the optimal throughput order $\frac{1}{\sqrt{n}}$ and the worst sensing distance scaling law, i.e., constant sensing distance, which is a special case in \eqref{eq:tradeoff}.
\demo
\end{remark}

\section{ISAC Networks with Fading}\label{sec:fading}
In this section, we address the problem of scaling law tradeoff in \emph{ad hoc} ISAC networks with random channel fading. We point out that the approaches from Sec. \ref{sec3} are not directly applicable here. Due to the added uncertainty, it is not clear whether random fading can be destructive to both communication and sensing SINR scaling and whether sufficient connectivity in the network exists to support a desired throughput under the constraints imposed by the sensing requirements.

\subsection{Channel Model}\label{sec:fading_model}
We consider the same network model as in Sec. \ref{sec:model}, except that all the transmitted signals suffer from small-scale fading in addition to path loss. Let $g_{k,j}$ denote the random fading gain for the communication link from node $X_k$ to node $X_j$. We assume that the fading is \emph{quasi-static} and \emph{independent} in each channel link. For $k,j \in \{1,\ldots,n\}$ and $k\neq j$, we assume that \cite{1354532}: 1) $g_{k,j}=g_{j,k}$; 2) the fading gain $g_{k,j}$ is real and nonnegative; 3) the expectation of $g_{k,j}$ is $\E[g_{k,j}]=1$; 4) the complementary cumulative distribution function (CCDF) of $g_{k,j}$,. i.e., $\bar{F}_g(x)$, has an exponentially decaying tail:
\begin{equation}\label{eq:ccdf}
\mathbb{P}[g_{k,j} > x] = \bar{F}_g(x)\leq q_0e^{-q_1x}, \forall x \geq g_0,
\end{equation}
for some real and positive parameters $q_0$, $q_1$, and $g_0$. We note that these assumptions are satisfied by most distributions used to model fading. Particularly, it has been proved in \cite[Lemma 5]{Toumpis03phd} that Rayleigh, Nakagami, and Ricean distributions all satisfy the properties in Item 4 above. It is also worth noting that the random fading gains corresponding to all the channels in the network can have a mixture of different distributions.

%
%


For communication, we still adopt the per-node throughput as the performance measure. Based on the fading model and follow \eqref{eq:comm_SINR}, the communication rate from node $X_k$ to its intended node $X_{R(k)}$ is given by
\begin{align}
R_k = \log_2\left(1+\frac{\frac{Pg_{k,R(k)}}{(r_{k})^{\alpha_{\text{c}}}}}{N_0+ \sum_{i\neq k, i\in \mathcal{T}}\frac{Pg_{i,R(k)}}{(d_{i,R(k)})^{\alpha_{\text{c}}}}}\right).
\end{align}

As for sensing, we consider that the radar transmitter only detects the target in its line-of-sight. Hence, the sensing channel gain is always 1. The sensing distance for node $X_k$ given the SINR constraints $\beta_{\text{s}}$ in \eqref{eq:sensing_SINR} satisfies
\begin{align}
d_{k} \leq 
(\beta_{\text{s}})^{-\frac{1}{\alpha_{\text{s}}}}\left(\frac{P\sigma^2_0}{N_0+ \sum_{i\neq k, i\in \mathcal{T}}\frac{Pg_{i,k}}{(d_{i,k})^{\alpha_{\text{c}}}}}\right)^{\frac{1}{\alpha_{\text{s}}}}.
\end{align}
Note that unlike the echo signal from the target, the interfering signals are in non-line-of-sigh.


In what follows, we present the achievable scheme for ISAC networks with random fading and the resultant tradeoff between throughput and sensing distance. However, we point out that proving the converse for this case is challenging due to the randomness introduced by fading, which will be addressed in our future work.

%
%
%

\subsection{Achievable Scheme}
We note that scaling the transmit power with $n$ is still necessary for improving the sensing distance scaling law in the presence of fading. Moreover, as shown in Sec. \ref{sec:alternative_scheme}, scaling the transmit power, the subsquare side length, and the number of transmission time slots differently is not beneficial to the scaling law tradeoff. Hence, we consider using the same function $f(n)$ to control those quantities as in Sec. \ref{sec3}.

\subsubsection{Bounding the Interference}
We first adopt the same setup as in Sec. \ref{sec:setup}, where square $S_n$ is partitioned into $45^{\circ}$-angled subsquares $s_i$ of a side length $\varsigma (f(n))^{\frac{1}{\alpha_{\text{c}}}}$, the communication distance satisfies $r_k \leq 2\sqrt{2}\varsigma (f(n))^{\frac{1}{\alpha_{\text{c}}}}$, and the number of transmission time slots $M>2$ is a constant. Again, we are only interested in the case of $f(n) = o(n^{\frac{\alpha_{\text{c}}}{2}})$, whereas the case of $f(n) = \Theta(n^{\frac{\alpha_{\text{c}}}{2}})$ is pure TDM, see Sec. \ref{Sec:TDM}. We have the following proposition that bounds the interference at each communication receiver and at each time slot w.h.p.

\begin{proposition}\label{prop2}
For $\alpha_{\text{c}}>2$ and all $k\in \mathcal{T}(m)$ and $m\in\{1,\ldots,M^2\}$, there exists a constant $\zeta>0$ such that the interference at node $X_{R(k)}$ satisfies
\begin{align}\label{eq:int_prop}
\lim_{n \rightarrow \infty}\mathbb{P}&\left[ \sum_{i\neq k, i\in \mathcal{T}(m)}\frac{Pg_{i,R(k)}}{(d_{i,R(k)})^{\alpha_{\text{c}}}}  \leq \zeta \log \frac{n}{(f(n))^{\frac{2}{\alpha_{\text{c}}}}},\forall k,m \right] \nonumber\\
&\rightarrow 1.
\end{align}
\end{proposition}
\begin{IEEEproof}
The proof is in Appendix \ref{app:proof_prop2}.
\end{IEEEproof}

The interference at each sensing receiver and at any time slot can be bounded similarly as in Proposition \ref{prop2}. Hence, we do not repeat the detailed proof here. It can be seen that the interference can no longer be upper bounded by a constant. This is because in the worst case, all fading gains are smaller than $\log n$ w.h.p. according to the CCDF in \eqref{eq:ccdf}, i.e., $\mathbb{P}\left[g \leq  \log n\right] \geq 1-q_0 n^{-q_1}$. Note also that the received signal power cannot be lower bounded by a constant due to random fading in general. At this point, it is not clear whether random fading could degrade both the communication and sensing performance of an \emph{ad hoc} ISAC network.

\subsubsection{Routing}\label{sec:routing_fading}
We follow the three-phase routing strategy from Sec. \ref{sec:routing} since it is beneficial to both communication and sensing in terms of achieving a constant data and a scalable sensing distance, respectively. However, it is important to realize that now the network connectivity depends on fading, in addition to whether there are nodes available for relaying the ISAC signals. Hence, we introduce new approaches to address this complication. In particular, we will show that by exploiting the spatial diversity \cite{1354532,4957662}, w.h.p. the information can be carried in the highway simultaneously across the network at a constant rate.

\begin{figure}[t!]
	\centering
\includegraphics[width=0.2\linewidth]{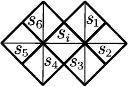}
\caption{Subsquare $s_i$ and its 6 neighboring subsquares whose edges are directly connected to that of $s_i$.}
\label{fig:subsquare}
\end{figure}

Following Sec. \ref{sec:routing}, we divided $S_n$ into subsquares $s_i$ of the same side length. Inspired by the sensing SINR threshold, we introduce the predetermined thresholds $g_{\tau}>0$ and $I_{\tau}>0$ for the channel gain and interference, respectively, for the highway phase. This is to ensure that the received signal power and interference meet their own threshold targets w.h.p. in the highway phase. Moreover, we will associate the connectivity, i.e., whether a subsquare and its edge are open or closed, with the thresholds. To this end, consider a horizontal subsquare $s_i$ with an edge across it horizontally. We note that there are 6 neighboring subsquares whose edges are directly connected to the edge of $s_i$, as illustrated in Fig. \ref{fig:subsquare}. With some abuse of notation, we let $g_{i,j}$ and $d_{i,j}$ represent the fading coefficient and distance, respectively, between a node in subsquare $s_i$ and a node in subsquare $s_j$. For any subsquare $s_i$ and its 6 neighboring subsquares $s_j$, $j=1,\ldots,6$, as  we declare that $s_i$ (and its edge) is \emph{closed} if at least one of the following conditions applies:
\begin{align}
&1) \quad\exists j\in \{1,2,3\}:g_{i,j}<g_{\tau}; \nonumber\\
&2) \quad\exists j\in\{1,\ldots,6\}:\sum_{k\neq j,k\in \mathcal{T}(m)}\frac{Pg_{k,j}}{(d_{k,j})^{\alpha_{\text{c}}}}>I_{\tau}. \nonumber
\end{align}
In other words, $s_i$ is \emph{open} if it contains at least one node and none of the above conditions apply. The first condition ensures that the channel gain from the node in the previous subsquare to the node in the next subsquare along the \emph{open} path is no less than $g_{\tau}$ (due to the reciprocity of the channel gain). The second condition ensures that the interference at each receiver node in any open neighboring subsquare of $s_i$ is no larger than $I_{\tau}$. The conditions for the status of vertical subsquares associated with the vertical open paths follow similarly. From the above two conditions, we see that a subsquare is open or closed is independent of the status of other subsquares in the network, allowing the mapping to the bond percolation model to facilitate the construction of the highway.

We now bound below the probability of each subsquare $s_i$ being open. Define the probability that a channel gain $g$ is no smaller than $g_{\tau}$ as
\begin{subequations}\label{eq:prop_g_tau}
\begin{align}
\mathbb{P}_g \triangleq& \mathbb{P}[g \geq g_{\tau}], \\
\geq& 1-e^{-q_3}, \label{eq:prop_g_tau2}
\end{align}
\end{subequations}
where \eqref{eq:prop_g_tau2} follows that there exists a constant $q_3>0$ such that $\mathbb{P}_g$ can be lower bounded by an exponentially decaying tail function according to Item 4 in Sec. \ref{sec:fading_model}. Then, we can immediately obtain the probability that the first condition does not occur is $(\mathbb{P}_g)^3$. Let $\mathbb{P}_I$ denote the probability that the second condition does not occur. We note that the mean of the interference satisfies
\begin{subequations}\label{eq:29}
\begin{align}
\mathbb{E}&\left[ \sum_{k\neq j,k\in \mathcal{T}(m)}\frac{Pg_{k,j}}{(d_{k,j})^{\alpha_{\text{c}}}}\right] \leq \varsigma^{-\alpha_{\text{c}}}\sum^{\infty}_{l=1}\sum^{8l}_{i=1}\frac{\mathbb{E}[g_{j+4l(l-1),j}]}{ (lM -2)^{\alpha_{\text{c}}}} \label{eq:29a}\\
=& \varsigma^{-\alpha_{\text{c}}}\sum^{\infty}_{l=1}\sum^{8l}_{i=1}\frac{1}{ (lM -2)^{\alpha_{\text{c}}}} \label{eq:29b} \\
=& \frac{8}{(\varsigma M)^{\alpha_{\text{c}}}}\sum^{\infty}_{l=1}\frac{l}{\left(l -\frac{2}{M} \right)^{\alpha_{\text{c}}}}<\infty, \label{eq:29c}
\end{align}
\end{subequations}
where \eqref{eq:29a} follows from \eqref{eq:int01} in Appendix \ref{app:proof_prop2}, \eqref{eq:29b} follows that $\mathbb{E}[g]=1$, and \eqref{eq:29c} follows from \eqref{eq:jl_int} in Appendix \ref{app:proof_prop2}. Then, the probability $\mathbb{P}_I$ can be lower bounded as follows
\begin{subequations}\label{eq:inf_tr}
\begin{align}
\mathbb{P}_I =& \prod^6_{j=1} \mathbb{P}\left[\sum_{k\neq j,k\in \mathcal{T}(m)}\frac{Pg_{k,j}}{(d_{k,j})^{\alpha_{\text{c}}}} \leq I_{\tau}\right] \label{eq:inf_tr0}\\
>& \prod^6_{j=1} \left(1-\frac{\mathbb{E}\left[ \sum_{k\neq j,k\in \mathcal{T}(m)}\frac{Pg_{k,j}}{(d_{k,j})^{\alpha_{\text{c}}}}\right]}{I_{\tau}}\right) \label{eq:inf_tr1}\\
\geq&\left(1-\frac{\frac{8}{(\varsigma M)^{\alpha_{\text{c}}}}\sum^{\infty}_{l=1}\frac{l}{\left(l -\frac{2}{M} \right)^{\alpha_{\text{c}}}}}{I_{\tau}} \right)^6, \label{eq:inf_tr2}
\end{align}
\end{subequations}
where \eqref{eq:inf_tr0} is due to the independence of the channel gain, \eqref{eq:inf_tr1} follows by using the Markov's Inequality $\mathbb{P}[X \geq a] \leq \frac{\mathbb{E}[X]}{a}$, and \eqref{eq:inf_tr2} follows from \eqref{eq:29}. Finally, combining \eqref{eq:p_open1}, \eqref{eq:prop_g_tau}, and \eqref{eq:inf_tr}, we obtain the probability of each subsquare $s_i$ being open as
\begin{equation}\label{eq:po_fading}
\mathbb{P}_{\text{o}} =\left(1-e^{-\varsigma^2 (f(n))^{\frac{2}{\alpha_{\text{c}}}}}\right)\left(\mathbb{P}_g\right)^3\mathbb{P}_I.
\end{equation}

We then incorporate the above threshold constraints in constructing the highway system of the network. To do so, we choose $g_{\tau}$ and $I_{\tau}$ such that
\begin{subequations}\label{eq:33}
\begin{align}
(\mathbb{P}_g)^3\mathbb{P}_I>&\frac{1-2e^{-\varsigma^2 (f(n))^{\frac{2}{\alpha_{\text{c}}}}}}{\left(1-e^{-\varsigma^2 (f(n))^{\frac{2}{\alpha_{\text{c}}}}}\right)^2} 
>\frac{1-2e^{-\varsigma^2 }}{\left(1-e^{-\varsigma^2 }\right)^2}, 
\end{align}
\end{subequations}
where \eqref{eq:33} follows that the function is increasing with $f(n)$ when $\varsigma f(n)>0$. By using the lower bound of $\mathbb{P}_{\text{o}}$ by substituting \eqref{eq:33} into \eqref{eq:po_fading}, one can follow the same proof as in Appendix \ref{sec:proof_th2} to show that Proposition \ref{th2} in Sec. \ref{sec:routing} still holds in this case. Hence, w.h.p. the highway system rich in crossing paths exists in the network. The rest of the routing steps follow those in Sec. \ref{sec:routing}.

\subsubsection{Tradeoff Between Throughput and Sensing Distance}
We now analyze the throughput and sensing distance for the ISAC network with random channel fading.

We start with the draining phase, where each source node communicates to an entry point node in the highway and performs sensing at the same time. By using the setup in Sec. \ref{sec:bound_acc}a, we know that the communication distance between the source node and the entry point node satisfies $r_k \leq \frac{\sqrt{2}\kappa\sqrt{n}}{\xi}\log \xi$. Moreover, the number of transmission time slots is given by $M^2=4(\sqrt{2}\kappa\log\xi+1)^2$. One can follow the proof of Proposition \ref{prop2} in Appendix \ref{app:proof_prop2} and show that w.h.p. the interference at each entry point node $X_j$ at any time slot $m$ is upper bounded as
\begin{align}\label{eq:35}
\sum_{i\neq k, i\in \mathcal{T}(m)}\frac{Pg_{i,j}}{(d_{i,j})^{\alpha_{\text{c}}}}  \leq \frac{\zeta \log \frac{n}{(f(n))^{\frac{1}{\alpha_{\text{c}}}}}}{(\sqrt{2}\kappa \log \xi+1)^{\alpha_{\text{c}}}} ,
\end{align}
for some constant $\zeta>0$. Next, we show that every source node in the ISAC network is able to choose a highway entry point such that the channel gain of the communication link between them is lower bounded by a constant w.h.p. Specifically,
\begin{subequations}
\begin{align}
&\mathbb{P}\left[\frac{Pg_{k,j}}{(r_k)^{\alpha_{\text{c}}}}>\frac{g_{\tau}}{\left(\sqrt{2}\kappa\varsigma \log \xi\right)^{\alpha_{\text{c}}}},\forall k\in\{1,\ldots,n\}\right] \nonumber \\
 =& \mathbb{P}\left[\max_{j\in\{1,\ldots,\eta \kappa \log \xi\}} \left\{g_{k,j}\right\}>g_{\tau},\forall k\in\{1,\ldots,n\}\right] \label{eq:36a} \\
=&\left(1-\left(\mathbb{P}[g_{k,j} \leq g_{\tau}]\right)^{\eta \kappa \log \xi} \right)^n \label{eq:36b} \\
\geq & \left(1-\left(\frac{\sqrt{n}}{\sqrt{2}\varsigma (f(n))^{\frac{1}{\alpha_{\text{c}}}}}\right)^{-\eta \kappa q_3}\right)^n,\label{eq:36c}
\end{align}
\end{subequations}
where \eqref{eq:36a} follows that a source node can choose one of the $\eta \kappa \log \xi$ highway entry point nodes whose channel gain is larger than $g_{\tau}$, \eqref{eq:36b} is due to the independence of the channel gain, and \eqref{eq:36c} follows from \eqref{eq:prop_g_tau} for some constant $q_3>0$. The RHS of \eqref{eq:36c} tends to 1 as $n \rightarrow \infty$ if we choose $\eta$ and $\kappa$ such that $\eta\kappa q_3 > \frac{\log n}{1/2\log n -1/\alpha_{\text{c}}\log f(n)-\log \sqrt{2}\varsigma}$. Therefore, w.h.p. the communication rate from the source to the entry point is
\begin{subequations}
\begin{align}
R_k = &\log_2\left(1+\frac{\frac{g_{\tau}}{\left(2\kappa\varsigma \log \xi\right)^{\alpha_{\text{c}}}}}{N_0+\frac{\zeta}{(\sqrt{2}\kappa \log \xi+1)^{\alpha_{\text{c}}}} \log \frac{n}{(f(n))^{\frac{1}{\alpha_{\text{c}}}}}}\right) \\
=&\Theta\left(\left(\log\frac{\sqrt{n}}{\sqrt{2}\varsigma (f(n))^{\frac{1}{\alpha_{\text{c}}}}}\right)^{-\alpha_{\text{c}}}\right).
\end{align}
\end{subequations}
Since the communication rate has the same order as that in \eqref{eq:draining_Rk}, we obtain the same per-node throughput order in the draining phase as in \eqref{eq:drain_t2}.

As for sensing, one can follow \eqref{eq:35} to show that the interference at each sensing receiver at any time slot of the draining phase is $\Theta\left(\left(\log \frac{\sqrt{n}}{f(n)^{1/\alpha_{\text{c}}}}\right)^{1-\alpha_{\text{c}}}\right)$ w.h.p. Hence, by \eqref{eq:dk_prop1} and \eqref{eq:dk1}, we obtain the same sensing distance order as in \eqref{eq:dk_no_I} in the draining phase.

In the highway phase, we apply the setup in Sec. \ref{sec:setup} and analyze throughput and sensing distance. To exploit the spatial diversity, we further divide each transmission time slot $m$ into 6 shorter time slots to allow each highway node to transmit to each of its 6 neighbor (see Fig. \ref{fig:subsquare}) once. Due to the introduced channel gain and interference thresholds in Sec. \ref{sec:routing_fading} and the existence of a sufficient number open paths as proved in Proposition \ref{th2}, we note that w.h.p. the communication rate between hops in each open path is a constant, i.e.,
\begin{equation}
R_k = \log_2\left(1+\frac{\frac{g_{\tau}}{(2\sqrt{2}\varsigma)^{\alpha_{\text{c}}}}}{N_0+I_{\tau}}\right).
\end{equation}
As a result, one can follow Sec. \ref{sec:bound_acc}b to show that the per-node throughput is the same as in \eqref{eq:13b}. Note that this is different from the case in the draining phase, where the interference cannot be upper bounded by a constant. This can be seen by noting from \eqref{eq:inf_tr} that the probability of interference being smaller than $I_{\tau}$ is a constant between 0 and 1, which cannot be made arbitrarily close to 1 as $n \rightarrow \infty$ in the draining phase due to the lack of sufficient spatial diversity.

Similarly, one can show that the interference at each sensing receiver at any time slot of the highway phase is upper bounded by $I_{\tau}$ w.h.p. In other words, constructing the highway is also beneficial to the sensing performance in terms of achieving bounded interference w.h.p. Consequently, the sensing distance scaling law is the same as that in \eqref{eq:dk2}.

The analysis of the delivering phase is very similar to that of the draining phase, which is omitted due to space limitations. By combining the results from all three phases, it follows that the tradeoff between throughput and sensing distance scaling laws is given by
\begin{align}\label{eq:tradeoff_fading}
(\lambda(n),d(n)) \hspace{-1mm}=\hspace{-1mm} \left(\Theta\left( \frac{1}{(f(n))^{\frac{1}{\alpha_{\text{c}}}}\sqrt{n}}\right),\Theta\left( (f(n))^{\frac{1}{\alpha_{\text{s}}}} \right)\right).
\end{align}
Comparing \eqref{eq:tradeoff} and \eqref{eq:tradeoff_fading}, we see that the scaling law tradeoff of \emph{ad hoc} ISAC networks is not affected by random fading.


\section{Numerical Results}

\begin{figure}[t!]
	\centering
\includegraphics[width=0.8\linewidth]{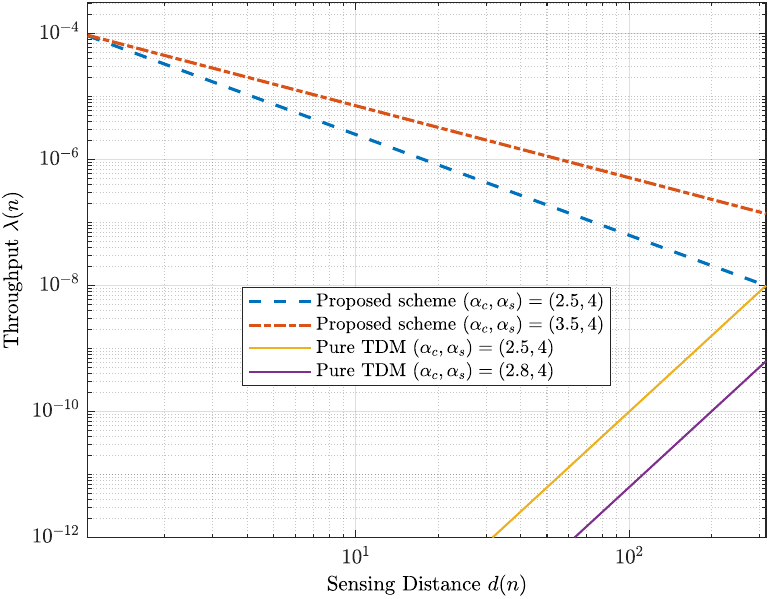}
\caption{Scaling law tradeoff of an \emph{ad hoc} ISAC network.}
\label{fig:result1}
\end{figure}

In this section, we compare the scaling law tradeoff of ISAC networks achieved by the proposed scheme and that by the benchmark pure TDM. We consider the node numbers as $n = 10^8$ let the transmit power be $P=f(n) = n^{\gamma}$, where $\gamma \in (0,1.25)$. Hence, all schemes use the \emph{same} transmit power. The throughput and scaling law tradeoff curves are obtained by varying $\gamma$ and shown in Fig. \ref{fig:result1}, where the path loss exponents $(\alpha_{\text{c}},\alpha_{\text{s}})$ are provided in the figure legend.

Surprisingly, the proposed scheme achieves a better scaling law tradeoff with a larger $\alpha_{\text{c}}$. This is because the proposed scheme reduces the hopping distance and allows more nodes to transmit ISAC signals simultaneously as the interference gets weaker with a larger $\alpha_{\text{c}}$. Consequently, the scaling law tradeoff can be improved. Meanwhile, the sensing distance order achieved by the proposed scheme does not change with $\alpha_{\text{c}}$. Intuitively, this can also be explained as follows: when fixing $\alpha_{\text{s}}$, the echo signal power of a sensing distance remains fixed. Then, with the larger $\alpha_{\text{c}}$, the interference power level is weaker, allowing more nodes to simultaneously transmit ISAC signals. Compared to the proposed scheme, the pure TDM scheme achieves inferior scaling law tradeoff due to the huge transmission time penalty on the throughput. Moreover, pure TDM is very sensitive to the path loss and its performance degrades as $\alpha_{\text{c}}$ increases.

In addition, we point out that unlike the link-level ISAC, obtaining a time-sharing scheme in ISAC networks is non-trivial as it requires an explicit design on transmission scheduling. For example, for $\delta,\tau\in (0,1)$, one can allow $n\delta$ nodes to perform communication-only tasks in $\tau$ time slot and $n(1-\delta)$ nodes to perform sensing-only tasks in the other $1-\tau$ time slots. However, the scheduling of these two sets of nodes to perform at two corner points, e.g., performing either at the communication-optimal point or sensing-optimal point, may not directly give another in-between tradeoff point for the scaling law.


\section{Conclusion}
In this paper, we investigated the tradeoff between throughput and sensing distance scaling laws in \emph{ad hoc} ISAC networks. We showed that scaling the transmit power according to the communication distance is the key to achieving the optimal scaling law tradeoff between throughput and sensing distance and provided achievability and converse proof. Specifically, our result reveals that by reducing the throughput with a function of $n$, the sensing distance increases with the same function of $n$ raised to the power of the ratio between the communication and sensing path loss exponents, regardless of with or without random fading. On the other hand, we showed that scaling the power and communication distance differently cannot improve the scaling law tradeoff. Our work opens up new research opportunities in the design of highly efficient large-scale ISAC networks.

For future work, it would be interesting to characterize the three-way tradeoff between communication, sensing distance, and sensing frequency of \emph{ad hoc} ISAC networks. In addition, the scheduling design for time-sharing schemes operating between the communication-optimal and sensing-optimal points in the network is another worthwhile direction.

\appendices
\section{Proof of Proposition \ref{th2}}\label{sec:proof_th2}
Let $\mathcal{A}_{R_{\text{h}},\eta \kappa \log\xi}$ be the event that $R_{\text{h}}$ has $\eta \kappa \log\xi$ disjoint open paths. Recall that each edge on $R_{\text{h}}$ is open with probability $\mathbb{P}_{\text{o}} $. Consider $\varsigma(f(n))^{\frac{1}{\alpha_{\text{c}}}} \geq 1$ and let $q = 1-2e^{-\varsigma^2 (f(n))^{2/\alpha_{\text{c}}}}$, such that $0<q<\mathbb{P}_{\text{o}} $ and $\frac{\mathbb{P}_{\text{o}} }{\mathbb{P}_{\text{o}} -q} = e^{\varsigma^2 (f(n))^{2/\alpha_{\text{c}}}}-1\leq e^{\varsigma^2 (f(n))^{2/\alpha_{\text{c}}}}$. By Lemma \ref{lem:percolation1} in Appendix \ref{sec:lemma}, we have
\begin{subequations}
\begin{align}
1 - \mathbb{P}[\mathcal{A}_{R_{\text{h}},\eta \kappa \log\xi}]  \leq& \frac{4}{3}\xi \left(e^{\varsigma^2 (f(n))^{\frac{2}{\alpha_{\text{c}}}}}\right)^{\eta \kappa \log \xi} \nonumber \\
&\cdot\left(6e^{-\varsigma^2 (f(n))^{\frac{2}{\alpha_{\text{c}}}}}\right)^{ \kappa \log \xi} \\
=&\frac{4}{3}\xi^{(\eta-1)\kappa \varsigma^2 (f(n))^{\frac{2}{\alpha_{\text{c}}}}+\kappa \log 6 +1}.
\end{align}
\end{subequations}

Since there are $\frac{\xi}{\kappa \log \xi}$ horizontal rectangles and the event $\mathcal{A}_{R_{\text{h}},\eta\kappa \log\xi}$ occurs independently in all $R_{\text{h}}$, the probability of \emph{each} $R_{\text{h}}$ having $\eta\kappa \log\xi$ disjoint open paths is given by
\begin{align}
(\mathbb{P}[&\mathcal{A}_{R_{\text{h}},\eta \kappa \log\xi}])^{\frac{\xi}{\kappa \log \xi}} \nonumber\\
&\geq \left(1-\frac{4}{3}\xi^{(\eta-1)\kappa \varsigma^2 (f(n))^{\frac{2}{\alpha_{\text{c}}}}+\kappa \log 6 +1}\right)^{\frac{\xi}{\kappa \log \xi}}. \label{eq:Anb1}
\end{align}
Notice that the RHS of \eqref{eq:Anb1} tends to 1 when $\xi \rightarrow \infty$ (as $n \rightarrow \infty$), provided that the following condition is satisfied
\begin{equation}
(\eta-1)\kappa \varsigma^2 (f(n))^{\frac{2}{\alpha_{\text{c}}}}+\kappa \log 6 +1  \leq -1.\label{eq:con1}
\end{equation}
It is easy to verify that the condition in \eqref{eq:con1} can be met as long as $\eta$ satisfies $0<\eta \leq 1-\frac{\kappa \log 6+2}{\kappa \varsigma^2 (f(n))^{\frac{2}{\alpha_{\text{c}}}}} \leq 1$ for any constant $\varsigma>0$. This completes the proof.

\section{Proof of Proposition \ref{prop2}}\label{app:proof_prop2}
As in Sec. \ref{sec:SINR}, we know that there are $8l$ interfering nodes on the $l$-th layer with distance at least $\varsigma(lM -2) (f(n))^{\frac{1}{\alpha_{\text{c}}}}$ from the receiver node $X_{R(k)}$, where $l=1,\ldots,l_{\max}$ and $l_{\max} = \left\lfloor\frac{\sqrt{n}}{\sqrt{2}\varsigma (f(n))^{\frac{1}{\alpha_{\text{c}}}}M} \right\rfloor$ is the maximum number of layers.
The interference at node $X_{R(k)}$ at time slot $m$ can be bounded as follows
\begin{subequations}\label{eq:int01}
\begin{align}
\sum_{i\neq k, i\in \mathcal{T}(m)}&\frac{Pg_{i,R(k)}}{(d_{i,R(k)})^{\alpha_{\text{c}}}} 
\leq\sum^{l_{\max}}_{l=1}\sum^{8l}_{j=1}\frac{Pg_{j+4l(l-1),R(k)}}{(\varsigma (f(n))^{\frac{1}{\alpha_{\text{c}}}}(lM -2))^{\alpha_{\text{c}}}}\\
=&\varsigma^{-\alpha_{\text{c}}}\sum^{l_{\max}}_{l=1}\sum^{8l}_{i=1}\frac{g_{j+4l(l-1),R(k)}}{ (lM -2)^{\alpha_{\text{c}}}}. \label{eq:int0}
\end{align}
\end{subequations}

Further to \eqref{eq:int01}, we then lower bound the following probability associated with the interference as
\begin{subequations}\label{eq:inf4}
\begin{align}
&\mathbb{P}\left[\varsigma^{-\alpha_{\text{c}}}\sum^{l_{\max}}_{l=1}\sum^{8l}_{i=1}\frac{g_{j+4l(l-1),R(k)}}{ (lM -2)^{\alpha_{\text{c}}}}  \leq  \zeta\log \frac{n}{(f(n))^\frac{1}{\alpha_{\text{c}}}} \right] \nonumber \\
=&\mathbb{P}\left[\varsigma^{-\alpha_{\text{c}}}\sum^{l_{\max}}_{l=1}\sum^{8l}_{i=1}\frac{g_{j+4l(l-1),R(k)}}{ (lM -2)^{\alpha_{\text{c}}}} \leq  \zeta'\sum^{l_{\max}}_{l=1}\sum^{8l}_{i=1}\frac{\log \frac{n}{(f(n))^\frac{1}{\alpha_{\text{c}}}}}{(lM -2)^{\alpha_{\text{c}}}} \right]\label{eq:inf1}\\
\geq &\prod^{l_{\max}}_{l=1}\prod^{8l}_{j=1} \mathbb{P}\left[g_{j+4l(l-1),R(k)} \leq \zeta' \log \frac{n}{(f(n))^\frac{1}{\alpha_{\text{c}}}}\right]\label{eq:inf2} \\
\geq & \left(1-q_0 \left(\frac{n}{(f(n))^\frac{1}{\alpha_{\text{c}}}}\right)^{-q_1 \zeta'}\right)^{4l_{\max}(l_{\max}+1)}\label{eq:inf3} ,
\end{align}
\end{subequations}
where in \eqref{eq:inf1} we define
\begin{align}\label{eq:zeta_prime}
\zeta'\triangleq \frac{\zeta}{\varsigma^{-\alpha_{\text{c}}}\sum^{l_{\max}}_{l=1}\sum^{8l}_{j=1}\frac{1}{(lM -2)^{\alpha_{\text{c}}}}},
\end{align}
which is upper bounded by some constant because the following sum
\begin{subequations}\label{eq:jl_int}
\begin{align}
\sum^{l_{\max}}_{l=1}\sum^{8l}_{j=1}\frac{1}{(lM -2)^{\alpha_{\text{c}}}} \leq& \sum^{\infty}_{l=1}\sum^{8l}_{j=1}\frac{1}{(lM -2)^{\alpha_{\text{c}}}} \\
=&\frac{8}{M^{\alpha_{\text{c}}}}\sum^{\infty}_{l=1}\frac{l}{\left(l -\frac{2}{M}\right)^{\alpha_{\text{c}}}},
\end{align}
\end{subequations}
is upper bounded by some constant due to $M>2$ and $\alpha_{\text{c}}>2$ as shown in \eqref{eq:IN1}, \eqref{eq:inf2} is due to the independence of each channel gain, and in \eqref{eq:inf3} we have used the probability bound in \eqref{eq:prob_bound} by using the CCDF of $g_{i,R(k)}$ in \eqref{eq:ccdf} for all $i\in\mathcal{T}(m)$
\begin{subequations}\label{eq:prob_bound}
\begin{align}
\mathbb{P}\left[g_{i,R(k)} \leq \zeta' \log x\right] \geq& 1-q_0e^{-q_1 \zeta' \log x} \\
=&1-q_0 x^{-q_1 \zeta'}.
\end{align}
\end{subequations}

%

With \eqref{eq:int01} and \eqref{eq:inf4}, we lower bound the probability in the LHS of \eqref{eq:int_prop} as
\begin{subequations}
\begin{align}
&\mathbb{P}\left[ \sum_{i\neq k, i\in \mathcal{T}(m)}\frac{P_ig_{i,R(k)}}{(d_{i,R(k)})^{\alpha_{\text{c}}}}  \leq \zeta \log \frac{n}{(f(n))^\frac{1}{\alpha_{\text{c}}}}, \forall k,m \right] \nonumber \\
\overset{\eqref{eq:int01}}{\geq}& \mathbb{P}\left[\varsigma^{-\alpha_{\text{c}}}\sum^{l_{\max}}_{l=1}\sum^{8l}_{j=1}\frac{g_{j+4l(l-1),R(k)}}{ (lM -2)^{\alpha_{\text{c}}}}  \leq  \zeta \log \frac{n}{(f(n))^\frac{1}{\alpha_{\text{c}}}}, \forall k,m \right] \\
\geq&  \Biggl(\mathbb{P}\Biggl[\varsigma^{-\alpha_{\text{c}}}\sum^{l_{\max}}_{l=1}\sum^{8l}_{j=1}\frac{g_{j+4l(l-1),R(k)}}{ (lM -2)^{\alpha_{\text{c}}}}   \nonumber \\
&\leq  \zeta\log \frac{n}{(f(n))^\frac{1}{\alpha_{\text{c}}}} \Biggr]\Biggr)^{M^2|\mathcal{T}(m)|} \label{ineq:inf_prop1} \\
\overset{\eqref{eq:inf3}}{\geq}& \left(1- q_0\left(\frac{n}{(f(n))^\frac{1}{\alpha_{\text{c}}}}\right)^{-q_1 \zeta'}\right)^{q_2\frac{n^2}{(f(n))^\frac{2}{\alpha_{\text{c}}}}},\label{ineq:inf_prop2}
\end{align}
\end{subequations}
where \eqref{ineq:inf_prop1} is again due to the independence of the channel gain at each active transmitter and at each time slot and \eqref{ineq:inf_prop2} follows from that $\lim_{n \rightarrow \infty}\frac{4l_{\max}(l_{\max}+1)}{n/((f(n))^{\frac{2}{\alpha_{\text{c}}}})} \leq q_2$ for some constant $q_2>0$ and $M^2|\mathcal{T}(m)| \leq n$ to ensure all nodes are considered. The RHS of \eqref{ineq:inf_prop2} tends to one as $n \rightarrow \infty$ if $\zeta$ is chosen to allow $\zeta'$ in \eqref{eq:zeta_prime} to satisfy $q_1 \zeta'>2$.

\section{Useful Lemmas}\label{sec:lemma}

\begin{lemma}\label{lem:percolation1}
Consider bond percolation in a rectangle $R$ containing $L_1\times L_2$ subsquares of a side length $\varsigma$, where each edge of the subsquares is open with probability $p$, independently of each other. For any $0\leq q< p \leq 1$ and $\eta>0$, the probability that there are at least $\eta L_2$ disjoint open paths that cross $R$ from left to right, i.e., $\mathbb{P}[\mathcal{A}_{R,\eta L_2}]$, satisfies
\begin{equation}
1-\mathbb{P}[\mathcal{A}_{R,\eta L_2}] \leq \frac{4}{3}m_1\left(\frac{p}{p-q}\right)^{\eta L_2 }(3(1-q))^{L_2}.
\end{equation}
\end{lemma}
\begin{IEEEproof}
The proof follows \cite[Appendix I]{4106120}.
\end{IEEEproof}

\begin{lemma}\label{lem:AN}
Consider that square $S_n$ is partitioned into $\frac{n}{\varsigma^2 (f(n))^\frac{2}{\alpha_{\text{c}}}}$ number of subsquare $s_i$ with a side length $\varsigma (f(n))^{\frac{1}{\alpha_{\text{c}}}}$. Then, there exists a constant $\delta>0$ such that each subsquare has less than $\varsigma f(n)^{\frac{1}{\alpha_{\text{c}}}}\sqrt{n} \left(\delta\log \frac{\sqrt{n}}{\sqrt{2}\varsigma f(n)^{\frac{1}{\alpha_{\text{c}}}}}\right)^{-\alpha_{\text{c}}-2}$ nodes with high probability.
\end{lemma}

\begin{IEEEproof}
Let $\mathcal{A}_n$ be the event that there is at least one subsquare $s_i$ containing more than $x$ nodes. Let $|s_i|$ denote the number of nodes in $s_i$. Then, we have
\begin{subequations}
\begin{align}
\mathbb{P}[\mathcal{A}_n]\leq&\frac{n}{\varsigma^2 (f(n))^{\frac{2}{\alpha_{\text{c}}}}}\mathbb{P}\left[|s_i|>x\right] \label{eq:AN1} \\
\leq&\frac{n}{\varsigma^2 (f(n))^{\frac{2}{\alpha_{\text{c}}}}}e^{-\varsigma^2 (f(n))^{\frac{2}{\alpha_{\text{c}}}}}\left( \frac{e \varsigma^2 (f(n))^{\frac{2}{\alpha_{\text{c}}}}}{x}\right)^{x}  \label{eq:AN2} \\
=& \left(\frac{e^{\frac{1}{x}\log n+2\log \varsigma (f(n))^{\frac{1}{\alpha_{\text{c}}}}+1}}{e^{\frac{1}{x}(\varsigma f(n))^{\frac{2}{\alpha_{\text{c}}}}+\frac{2}{x}\log\varsigma f(n)^{\frac{1}{\alpha_{\text{c}}}}+\log x}}\right)^x,
\end{align}
\end{subequations}
where \eqref{eq:AN1} follows the union bound and \eqref{eq:AN2} follows \cite[Eq. (17)]{4106120}, with the condition that $x>\varsigma^2 (f(n))^{\frac{2}{\alpha_{\text{c}}}}$. We know $x$ must be an increasing function of $n$ since more nodes are inside $s_i$ as its area becomes large. Let $<_a$ denote that the corresponding inequality will only hold \emph{asymptotically} for sufficiently large $n$, e.g., $f(n)<_a g(n)$ means that there exist $n_0$ such that $f(n)<g(n),\forall n>n_0$. As a result, the necessary condition for $\mathbb{P}[\mathcal{A}_n] \rightarrow 0$ as $n \rightarrow \infty$ is given by
\begin{align}
&\frac{\log n}{x}+2\log \varsigma (f(n))^{\frac{1}{\alpha_{\text{c}}}}+1 <_a\frac{\varsigma^2 (f(n))^{\frac{2}{\alpha_{\text{c}}}}}{x}\nonumber\\
&+\frac{2}{x}\log\varsigma (f(n))^{\frac{1}{\alpha_{\text{c}}}}+\log x  \\
\Rightarrow & \log n - \varsigma^2 (f(n))^{\frac{2}{\alpha_{\text{c}}}}+2x\log \varsigma (f(n))^{\frac{1}{\alpha_{\text{c}}}} -x\log x <_a 0. \label{ineq:logxn}
\end{align}
To fulfill the condition in \eqref{ineq:logxn}, we let
\begin{align}
x =&\varsigma^2 f(n)^{\frac{2}{\alpha_{\text{c}}}}x'\label{eq:x_lem}\\
x'=&\frac{\sqrt{n}}{\varsigma f(n)^{\frac{1}{\alpha_{\text{c}}}}} \left(\delta\log \frac{\sqrt{n}}{\sqrt{2}\varsigma f(n)^{\frac{1}{\alpha_{\text{c}}}}}\right)^{-\alpha_{\text{c}}-2}, \label{eq:x_lem1}
\end{align}
for some constant $\delta>0$. Clearly, $x'$ is an increasing function of $n$. Moreover, one can pick $\delta$ small enough to ensure $x'>1$ to meet the condition $x>\varsigma^2 (f(n))^{\frac{2}{\alpha_{\text{c}}}}$. Substituting \eqref{eq:x_lem} into the LHS of \eqref{ineq:logxn}, we get
\begin{subequations}
\begin{align}
&\log n - \varsigma^2 (f(n))^{\frac{2}{\alpha_{\text{c}}}} -\varsigma^2 f(n)^{\frac{2}{\alpha_{\text{c}}}}x'\log x' \nonumber \\
<&\log n - \varsigma^2 f(n)^{\frac{2}{\alpha_{\text{c}}}}x' \\
\overset{\eqref{eq:x_lem1}}{=}&\frac{\log n \left(\delta\log \frac{\sqrt{n}}{\sqrt{2}\varsigma (f(n))^{\frac{1}{\alpha_{\text{c}}}}}\right)^{\alpha_{\text{c}}+2}- \varsigma (f(n))^{\frac{1}{\alpha_{\text{c}}}}\sqrt{n}}{\left(\delta\log \frac{\sqrt{n}}{\sqrt{2}\varsigma (f(n))^{\frac{1}{\alpha_{\text{c}}}}}\right)^{\alpha_{\text{c}}+2}} \\
<_a& 0, \label{eq:node_num1}
\end{align}
\end{subequations}
where \eqref{eq:node_num1} follows from $(\log n)^{\alpha_{\text{c}}+3}<_a \sqrt{n}$. This completes the proof.
\end{IEEEproof}

\bibliographystyle{IEEEtran}
\bibliography{MinQiu}

\end{document}